\documentstyle[tighten,prd,aps]{revtex}
\begin{document}
\draft
\title{A model of quark and lepton masses I: The neutrino sector} 
\author{P. Q. Hung\cite{email}}
\address{Dept. of Physics, University of Virginia, \\
382 McCormick Road, P. O. Box 400714, Charlottesville, Virginia 22904-4714}
\date{March 30, 2000}
\maketitle
\begin{abstract}
If neutrinos have masses, why are they so tiny? Are these masses of the Dirac 
type or of the Majorana type? We are already familiar with the mechanism of 
how to obtain a tiny Majorana neutrino mass by the famous see-saw
mechanism. The question is: Can one build a model in which a tiny
{\em Dirac} neutrino mass arises in a more or less ``natural'' way?
What would be the {\em phenomenological consequences} of such a scenario,
other than just merely reproducing the neutrino mass patterns
for the oscillation data?
In this article, a systematic 
and detailed analysis of a model is presented, with, as key components, 
the introduction of a family symmetry as well as a new 
$SU(2)$ symmetry for the right-handed neutrinos. In particular,
in addition to the calculations of light neutrino Dirac masses, interesting
phenomenological implications of the model will be presented.
\end{abstract}

\section{Introduction}

There are strong indications- the latest of which came from the SuperKamiokande
collaboration \cite{superk}- that neutrinos do have a mass, albeit a very tiny one,
and, as a result, ``oscillate''. The exact nature of the masses as well as the
oscillation angles is an important subject which is under intense investigation
\cite{kayser}. 
Consequently, there exists many interesting models which, in one way or
another, try to accomodate most of the known data. It is perhaps
prudent to think that the subject of neutrino masses and oscillation is still
a very open one.

It is fair to say that the extreme smallness of neutrino masses suggests
something very peculiar about these particles. This peculiarity could
come from the way the neutrinos obtain their masses and/or from the
very special nature of the neutrinos themselves which distinguish them
from all other particles. For example, do right-handed neutrinos (present
in most models of neutrino masses) carry quantum numbers which 
are absent in some or all
other (left- or right-handed) fermions? After all, right-handed
neutrinos, if present, would be singlets under $SU(3) \otimes
SU(2)_L \otimes U(1)_Y$ anyway. 

Most efforts on the problem of neutrino masses, at least on the model-building
front, are concentrated on the construction of lepton 
mass matrices based on various ansatzes.
There is one common assumption present in many of such models, which is
one in which light neutrino masses arise from a see-saw mechanism \cite{seesaw}. 
The smallness
of neutrino masses would come from an expression that goes like $m_D^2/\cal{M}$,
where $m_D$ is a Dirac mass , and $\cal{M}$ is a Majorana mass 
which typically is very much larger
than $m_D$ . In these models, the scale
of new physics $\cal{M}$, as suggested by the lightness of neutrino masses, 
would be some kind of Grand Unified scale or even the breaking scale of
Left-Right symmetry models \cite{mohapatra1}. (Lepton number is not
a conserved quantity in this class of models.)
The see-saw mechanism is a very elegant
approach which is widely embraced. 

However, one could not help but
wonder if there might be some other mechanism for obtaining
tiny neutrino masses, and if so, how
it would fare compared with the see-saw mechanism. Would this new
mechanism shed light on other important issues?
What would be its
scale of new physics? 
Can one find an experimental distinction between the two mechanisms? 
This was the topic discussed in \cite{hung}. 

At the present time, it is not clear that, if neutrinos do 
have a mass, it would be of the Majorana or Dirac type. As we have mentioned
above, with Majorana neutrinos and the see-saw mechanism, one could
``easily'' obtain small neutrino masses. Now if the mass were to be of the
Dirac type, one can straightforwardly write down a gauge-invariant Yukawa coupling in the
SM itself (endowed with right-handed neutrinos, of course). But to obtain
a small neutrino mass, one has to put in {\em by hand} a Yukawa coupling which
is incredibly small, of the order of $10^{-11}$. Such a fine tuning is highly
unnatural and that might be the reason why little attention is given to
the construction of models based on Dirac neutrino masses. Did we leave
something out by ignoring it? What if the mass is truly of the Dirac type?
Until this question is settled, it is worthwhile to investigate possible
alternatives to the see-saw mechanism.
This paper and a previous one \cite{hung} propose one of such alternatives by
constructing a model of {\em Dirac} neutrino masses where the
smallness of their values arises dynamically. One of the criteria
used in building such a model is the wish to go beyond the mere
presentation of a neutrino mass matrix. In particular, we would
like to see if there might be other {\em phenomenological
consequences} which could be testable: New particles, new
physics signals, etc.. This is the aim we set about in building
our model.

The construction of the model presented in \cite{hung} was based on
the following questions: If neutrino masses were so small compared
to all other known masses, would there
be an appearance of a special symmetry when one lets the mass go to zero?
Could this special symmetry, if it exists, be a peculiar feature of the
right-handed neutral leptons alone?
Could there be additional purposes for its
existence other than providing a small mass for the neutrinos? In other words, can
one learn something more from it? It was found in Ref. \cite{hung} that
there is indeed an interesting symmetry which acts only on the right-handed
neutrinos and which, in addition to providing a reason for the smallness
of the neutrino masses, also constrains the nature (even or odd) of the
number of generations. Furthermore, the way in which neutrino masses are
constructed can be used to build a model for charged lepton 
and quark masses. In addition, this particular
way of constructing masses might even have some bearing on the strong CP problem.
Last but not least: Are there additional tests of various neutrino models
other than neutrino oscillations? For the see-saw mechanism with
Majorana neutrinos, one already sees that one of such additional signals is,
for example, the phenomenon of neutrinoless double beta decay. As it will
be presented below, the addtional signals of the model presented here will
involve a number of very concrete predictions: the absence of neutrinoless
double beta decay, the possible presence of ``low mass'' ( a couple
of hundreds of GeV e.g.) vector-like fermions, among other things. In particular,
the detection of these vector-like fermions do not in any way involve neutrinos.

One particularly important feature of our model is the following predictions
for neutrino oscillations, assuming only the validity of the atmostpheric and
solar neutrino data: 1) The three light neutrinos are nearly degenerate;
2) If the light neutrinos have a mass large enough to form a component of
the Hot Dark Matter (HDM) \cite{caldwell} then {\em only} the 
MSW solution to the solar neutrino
oscillation is favored; 3) If the vacuum solution to the solar neutrino problem
turns out to be the correct one, our model will only be able to
accomodate tiny neutrino masses, around $10^{-3} eV$ or less, ruling
out near-degenerate neutrinos as components of HDM. As a result, in our model,
one cannot have both vacuum solution and HDM. We will show below the correlation
between the masses and the differences of mass squared,
$\Delta m^2$, which enter
the neutrino oscillation phenomena.


Assuming the existence of the aforementioned symmetry, how can one 
construct Dirac neutrino masses  to be dynamically
small? By ``dynamically'', it is meant that the mass is
zero at tree level and that any non-zero value would have to arise at the
one-loop (or more) level. 
Now, the peculiar (and toughest) thing about neutrinos is the fact 
that their mass
is so small- at least eleven orders of magnitude smaller than the electroweak
scale. In constructing our model for Dirac neutrino masses, it is then reasonable to
ask under what conditions would the dynamical Dirac mass of the neutrinos obtained at the
one loop level be ``naturally'' small, i.e. devoid of excessive fine tuning.
In this paper, we present the following interesting results: In the four-generation 
model, it is found that the fourth neutrino can be naturally heavy while the
other three obtain their masses at one loop, with the result that these masses
can be tiny provided  some ratios of masses of particles which participate in the
loop integration are ``large'', {\em regardless} of their actual values. 
This is interesting because, as we shall see below, some of the particles
which participate in the loop integration, in particular the lightest ones,
can have masses as low as a few hundred GeVs and which could provide
a direct test of this model. We will also see that, in order to obtain very
small neutrino masses, at least one of the particles needs to be much
heavier than the lightest one- a result which is somewhat reminescent 
of the see-saw mechanism. 
We will also see that the mass of the light neutrinos is intrinsically tied
to the extra global symmetry present in the scalar sector of the model. In fact,
the extra Nambu-Goldstone (NG) bosons which are not absorbed by the (family and
$SU(2)_{\nu_R}$) gauge bosons acquire a mass due to the presence of the
gauge-invariant ``cross-coupling'' terms in the potential 
which explicitely break the extra
global symmetry. 

The above  brief statement will be made clearer in the discussion of 
neutrino masses. Notice, in particular, that the result given 
for light neutrino masses in \cite{hung} 
is only a very special case of the present discussion.

The plan of the paper is as follows. First, the model is presented with
a description of the gauge structure along with its particle content.
It is shown how a new symmetry prevents neutrinos from obtaining a
mass unless it is broken.
Next, the special properties of this extra symmetry associated with the
right-handed neutrinos are discussed. In particular, if that symmetry is
a chiral $SU(2)$ as is the case in this paper, nontrivial constraints
coming from the nonperturbative Witten anomaly \cite{witten}
can be applied to the
nature of the number of families. This is the extra bonus mentioned
above. The paper then proceeds to discuss the generation of light 
neutrino masses, principally by radiative corrections of the type
mentioned above. It is then followed by a
discussion of the neutrino mass matrix. In particular, we will present
the correlation between the values of the neutrino masses and $\Delta m^2$.
Most importantly, we will show how  $\Delta m^2$ increases or decreases with
the masses themselves, with two resulting implications: either
one has HDM {\em and} MSW or vacuum solution and {\em no} HDM. Either of
these solutions will have an important cosmological implication.
We end the paper with a brief discussion of the charged lepton mass
matrix, the primary purpose of which being the wish to complete the
discussion by presenting some examples on what the oscillation
angles might look like. A followed-up paper will deal seperately
with the charged lepton sector and, as a consequence, with a full
discussion of the angles.

We would like to emphasize for the purpose of clarity that the charged
lepton sector (which will be dealt with in a separate paper) is different
in structure from the neutrino sector, as we shall see below, and
does not have the same hierarchical structure.
The fact that, in this model, the three light neutrinos are nearly
degenerate does not imply that it would be the same in the charged
lepton sector. In fact, it is not as we will show in a subsequent
paper.

Finally, a section will be
devoted to various other phenomenological implications of the model.

We shall assume throughout this paper the existence of right-handed
neutrinos.

Since this manuscript is meant to be comprehensive, and hence
lengthy, one could skip the three subsections of the next section,
after first reading its introduction.(Its reading is nevertheless
recommended because the physics motivations are discussed there.)

\section{A Model}

It is well-known that all that is needed to give neutrinos a mass is
to simply add extra right-handed neutrinos to the Standard Model. One
can then construct a (Dirac) mass term with an arbitrary Yukawa coupling,
$g_{\nu} \bar{l}_L \phi \nu_R + H.c.$, 
which can be made to be as small as one wishes. This, of course, is
unsatisfactory because, if neutrinos have masses in the $eV$ range or less, 
this would require the Yukawa coupling, $g_{\nu}$, to be of O($10^{-11}$) 
or less. Fine tuning to such a precision is normally considered to be unnatural.
At this point, one might be tempted to try to explain this fact
by simply invoking a fourth generation with a democratic mass matrix, at
least for the neutrinos, as has been done by Ref. \cite{silva-marcos}.
The diagonalization of the neutrino mass matrix would then give
one heavy eigenstate and three massless states. By adding some arbitrary
phases to the mass matrix, one can ``provide'' a small mass (
depending on the values of those phases) to the three neutrinos. This
purely phenomenological ansatz (Ref. \cite{silva-marcos}) appears
to ``fit'' the recent data on neutrino oscillations with the
appropriate {\em choices} of the phases. However, the fourth generation
lepton masses came out to be extremely heavy and split, which practically
seems to be ruled out by analyses of precision experiments
\cite{erler}.

In \cite{hung}, a model of Dirac neutrino masses was constructed and based on 
a four generation scenario that was very different from the democratic
ansatz made in \cite{silva-marcos}. One of the reasons for
using such a scenario is the fact that, as of the present time,
a fourth generation is {\em not} ruled out by experiment and,
as a consequence, it is interesting to explore its possible implications.
A recent review \cite{hung2} gave a comprehensive discussion of
various topics concerning quarks and leptons beyond the third 
generation, including the present experimental status and
future searches.

If a fourth generation were to be used in the investigation of neutrino
masses, one should keep in mind various phenomenological constraints
concerning not only leptons but also quarks. For instance,
constraints on the $\rho$ parameter limit the mass splitting
within each doublet of extra quarks and leptons: the up and down
members of a fourth generation should be very close in mass. They
should be long-lived enough to escape present detection. This, in
turns, tells us something about the mixing between the fourth
generation and the other three. All of these issues have been
described in \cite{hung2}. In the construction of
the model presented in Ref. \cite{hung}, these phenomenological
constraints were kept in mind.

As mentioned briefly in the Introduction, our approach, as described
in Ref. \cite{hung}, is based on a dynamical justification for
for the small value of the neutrino Yukawa couplings.
The question that was asked was: Could there be a scenario in
which a symmetry appears as one lets the Yukawa coupling go to
zero? The tiny Yukawa coupling which would give the neutrino a
very small mass would then arise dynamically when that symmetry is broken.
These Yukawa couplings then appear as effective couplings which could
be small for dynamical reasons and are not
fundamental parameters that are put in by hand and which are needed to
be fine tuned.
What is the nature of that symmetry and how a dynamical Yukawa coupling
appears will be the subject of this section and the following two.

It is obvious that an extension of the Standard Model(SM) is needed
in addressing the above issues. One simply cannot stay solely within the
SM if one wishes to deal with the mass of the neutrinos. What it is that
one needs when one goes beyond the SM is a matter of taste, modulo
a very obvious requirement: predictability of new phenomena or particles
which can be tested.

We first describe the model, presenting its gauge structure and representations.
Next, explanations are provided for the reasons behind the choices
of the extended gauge group and its particle content. The crucial
assumption here is the existence of two new symmetries, one of which
will be particular to right-handed neutrinos, as alluded to
earlier, and the other one is a family gauge symmetry. As we shall see
below, it is the breaking of these new symmetries that will give
a mass to the neutrinos. 

In this work, the SM is extended in the following way. Generically,
it takes the form: $SU(3)_c \otimes SU(2)_L \otimes U(1)_Y \otimes
(Family\, symmetry) \otimes (right-handed\, neutrino\, special\, symmetry)$.
Why a ``Family symmetry''? This is so for two reasons: a) We wish
to investigate the family replication problem and the mixing among
different generations; b) The special symmetry endowed by the
right-handed neutrinos might have some bearing on the family
symmetry itself. After all, if one would like to investigate
the family problem, some kind of family symmetry-be it discrete
or continuous, global or gauge- is needed.
Why a special symmetry for the right-handed
neutrinos? The reasons were already expounded above: To provide
a framework for an understanding of the smallness of neutrino
masses. Our next task is then to determine what this special
symmetry might be and what form the family symmetry might take.

Our model is described by:
\begin{equation}
SU(3)_c \otimes SU(2)_L \otimes U(1)_Y \otimes SO(N_f) \otimes SU(2)_{\nu_R}
\end{equation}
where $SO(N_f)$ and $SU(2)_{\nu_R}$ are the family gauge group and the
special gauge group for the right-handed neutrinos respectively. The
particle content of the model is listed in Table 1. Notice that we have
denoted the right-handed neutrinos by
$\eta_R =(\nu_R^{\alpha}, \tilde{\nu}_R^{\alpha})$ because they are
assumed to transform as doublets under $SU(2)_{\nu_R}$. The two options
listed for the right-handed neutrinos as well as the meaning of the
non-standard particles will be discussed below. We would first like to 
explain the choices of the extra gauge groups. Here,
the extra symmetries are chosen to be {\em gauge symmetries}
because, as it is well known, powerful constraints can be obtained from
models built on the gauge assumption.

\subsection{Why $SU(2)_{\nu_R}$?}

Let us first look at Table 1. 
In this model, all standard (left-handed and right-handed)
particles are singlets under $SU(2)_{\nu_R}$. Hence the subscript
$\nu_R$. In this respect, $SU(2)_{\nu_R}$ is very different from
$SU(2)_R$ of the popular Left-Right model \cite{mohapatra1}. 
In that model, right-handed
quarks and leptons form doublets under $SU(2)_R$, {\em for every family}.
Because of our assignment, all weak interactions among standard
particles are pure V-A, in contrast with the Left-Right model.
What is the motivation behind our choice that makes it so
different from the Left-Right model? To answer that question,
let us recall an interesting feature of chiral $SU(2)$: the
presence or absence of the so-called Witten global anomaly.

If chiral fermions transform as doublets under $SU(2)$, there exists
a nonperturbative anomaly- the so-called Witten anomaly \cite{witten}- 
associated with an {\em odd} number of doublets. Briefly
speaking, this is so because the fermionic determinant
$\sqrt{det\, i \not\!\nabla (A_{\mu})}$ changes sign under
a ``large'' gauge transformation $A_{\mu}^{U} = U^{-1}
A_{\mu} U - i U^{-1} \partial_{\mu} U$
if the number of chiral doublets is odd. This would make the partition
function $Z$ vanish and the theory would be ill-defined. This
nonperturbative anomaly would then require the number
of Weyl doublets to be {\em even} in order for the theory to be consistent. 
(This ambiguity in sign stems from the fact that the fourth homotopy
group $\Pi_4 (SU(2)) = Z_2$.) Other groups that also have similar
non-trivial constraints are $Sp(N)$ for any $N$ and $O(N)$ for
$N \leq 5$.

It is amusing to recall a well-known but forgotten fact about the SM. There
the chiral gauge group is $SU(2)_L$. Each family contains one
lepton and three quark doublets and, as such, is free from
the global Witten anomaly. (Let us recall that the cancellation
of the {\em perturbative} triangle anomaly in the SM only relates
the lepton charge to that of the quark.)  If, instead of three,
the number of colors, $N_c$, were arbitrary, the freedom from such an 
anomaly would require $1+N_c$ to be
even, and hence, $N_c$ to be odd, namely $N_c$ = 3, 5,...
Why nature choses $N_c =3$ instead of some other
odd number is a question which can only be answered in the context of some
deeper theory such as e.g. $SU(5)$. Although the Witten 
anomaly does not fix the size
of $N_c$, it is nevertheless a powerful constraint 
in the sense that, {\em once
a fermion content is known} (e.g. one color singlet (leptons) and one
fundamental representation (quarks) in the SM), $N_c$ is constrained
(e.g. odd in the case of the SM). 

The above simple lesson taught us something about the powerful constraint
that a chiral $SU(2)$ exerts on the number of chiral doublets. This is
the reason why it is chosen to be the special symmetry of the right-handed
neutrinos. Let us contrast the constraint coming from $SU(2)_{\nu_R}$
with that coming from $SU(2)_R$ (Left-Right model). For our model,
with $SU(2)_{\nu_R}$, {\em only} $\eta_R$ transforms as doublets. Absence
from the Witten anomaly then requires the number of such doublets to be
{\em even}. If $\eta_R$ carries, in addition, family indices then the
anomaly requirement restricts the number of generations to be {\em even}
such as in Option 1 as indicated in Table 1. If there exists an $\eta_R$ which
is a family singlet (denoted by $\eta_R^{\prime}$), the number of generations
would be {\em odd} such as in Option 2 of Table 1. With the Left-Right model,
{\em each family} contains four doublets of $SU(2)_R$: $(\nu_R, e_R)$
and $(u_R, d_R)_{i}$ with $i=1,..,3$. Therefore, the Witten 
anomaly requirement is automatically
satisfied {\em per} family. This is one of the few differences between our
model and the Left-Right model. 

A final word of caution is in order here. Although the Witten anomaly
constraint allows us to make a statement on the evenness or oddness
of the number of generations- a subject to which we shall come back
in the next subsection, it {\em does not} determine that number. This
should come from a deeper and as-yet-unknown theory. Our goal is much
more modest: Given a fermion content (Option 1 or 2 below), we can say whether
or not the number of generations is odd or even, and that is all. We shall
however try to constraint that number from a different route which is more
phenomenological in nature, and point out the differences between
Option 1 and 2.

\subsection{Why $SO(N_f)$?}

In the construction of any model, there is a time-honored requirement:
the absence of the perturbative triangle anomaly. Even if the Witten
anomaly were absent, this requirement is a must for any gauge theory.
(It just happens that,in the SM, both requirements are simultaneously
satisfied.) In our case, if a family index is assigned to
all standard fermions {\em and} to $\eta_R$, the family gauge group
that is chosen cannot be a vector-like theory, which is anomaly-free,
because $\eta_R$
posseses an additional quantum number, that of $SU(2)_{\nu_R}$. This
is unlike QCD or even the Left-Right model if left and right-handed
fermions carry similar family quantum numbers. A safe group and
representations have to be chosen.

The choice made in this paper is $SO(N_f)$ for the family gauge group,
with chiral (left- and right-handed) fermions transforming as (real)
vector representations with $N_f$ components each. As such, the model
is also free of the perturbative triangle anomaly.

Our model based on $SU(3)_c \otimes SU(2)_L \otimes U(1)_Y \otimes 
SO(N_f) \otimes SU(2)_{\nu_R}$ with an {\em even} number of $SU(2)_{\nu_R}$-
doublets and chiral fermions transforming as {\em vector representations}
of $SO(N_f)$ is free from both nonperturbative and perturbative
anomalies.

\subsection{Constraints on $N_f$}

As shown in Table 1, there are two options for $\eta_R$, each of which
should contain an even number of $SU(2)_{\nu_R}$ doublets.

a) Option 1:

$\eta_R^{\alpha}$ carries the family index $\alpha = 1..N_f$ where
$N_f= 2,4,6, 8,..$.

b) Option 2:

Here we have $\eta_R^{\prime}$ (a family singlet) and $\eta_R^{\alpha}$.
The constraint is now $1 + N_f =$ even, which means that $N_f = 3, 5, 7,..$
(excluding the trivial case of 1 family).

Unlike the SM where one knows the fermion content for each family, i.e.  quarks and leptons, 
and hence the nature of $N_c$- it is {\em odd}- our scenario involves
incomplete experimental informations, and as such, the nature (odd or even) of
$N_f$ cannot be completely fixed. Each choice, however, represents a distinct
particle content (no family singlets for the even option and one family singlet
for the odd option) which implies a possible distinct route for a yet-unknown 
unification. 

Recognizing the fact that there are deep differences between the even and
odd options-a point to be discussed below- and in the absence of a deeper
theory, one might wonder what can be done to narrow down the choices, not between
odd or even, but within each option itself. Below we present an argument
that could help in finding a way to further restrict $N_f$. This argument
is only suggestive, being a combination of ``theoretical prejudice''
and phenomenological constraint.

One might require that gauge couplings are free from
Landau singularities below the Planck scale in such a way that
unification of the SM gauge couplings, if it exists, occurs in the 
perturbative regime \cite{hung1}.  
With this criterion, one can see that the even
option can only accomodate $N_f = 2, 4, 6$, while the odd option can only 
accomodate $N_f = 3, 5$. This is because for $N_f \geq 7$, one or more
gauge couplings will ``blow up'' before the Planck scale. There are
no reasons, in the absence of a deeper theory, to rule out any of 
the above choices. This will require other yet-unknown conditions.
The only thing one can say, in the context of our model, is that
electroweak precision experiments appear to
rule out $N_f \geq 5$ \cite{erler} and and that existential facts tell 
us that $N_f$ is at
least three. This leaves us with the choice $N_f=4$ for the even option
and $N_f = 3$ for the odd option.

If $N_f \leq 4$ comes from the above argument, what then is
the role of the Witten anomaly in all of this? It tells us about the
particle content of the right-handed neutrinos. For $N_f$ = 4, the
right-handed neutrinos are simply $\eta_R = (1, 1, 0, 4, 2)$ while
for $N_f$ = 3, one has $\eta_R = (1, 1, 0, 3, 2)$ plus a family
singlet $\eta_R^{\prime} = (1, 1, 0, 1, 2)$. What observed differences
can there be between these two options? 
The former predicts the existence of a fourth generation whose consequences
have been recently discussed in Refs. \cite{hung1} and  \cite{hung2,physrep}. The
latter predicts the existence of a neutral family-singlet $\eta^{\prime}_R$
(doublet under $SU(2)_{\nu_R}$) which could have cosmological
consequences of a yet-unknown nature. In addition, 
as we point out below, it appears that the
even option prefers three almost degenerate light neutrinos while the odd
option prefers a hierarchical structure for the light neutrinos.   
If a fourth generation is discovered, which alone does not necessarily 
imply the even option presented here,
and if the light neutrino masses are convincingly ``proven'' to
be nearly degenerate (instead of a hierarchical structure), the even option
might be viable. Furthermore, as we shall see below, another
possibility for testing this model is to look for signals of some of the
lightest particles- the vector-like fermions- which participate in
the loop diagram of Fig. 1. As discussed below, the light neutrino masses
depend only on the ratios of these masses and not on their magnitudes
and these vector-like fermions can be as light as a few hundred
GeVs.

\section{Neutrino Masses}

This section will be devoted to the discussion of how neutrino masses can be
generated in our model for Option 1. We shall comment on Option 2
at the end of the manuscript.
We shall concentrate only on the lepton sector and,
in particular, on the neutrino one, leaving the full discussion
of the charged lepton and
quark sectors for a subsequent publication.

Since we will be dealing only with {\em Dirac} neutrino masses, we 
shall require that {\em all}
fermions be endowed with a global
$B-L$ symmetry. Since we are concerned only with leptons in this
section, a global $L$ symmetry is sufficient for the present
purpose. This global $L$
symmetry would {\em prevent} a Majorana mass term of the type
$\eta^{i\,\alpha}_{R} \eta_{i\,\alpha\,R}$, where $i=1,2$ and
$\alpha = 1,..,4$. {\em Only Dirac masses
will be allowed}. 

There might be other suggestive reasons as to why Dirac masses 
for the neutrinos might be 
attractive. For example, a combined fit of massive neutrinos
as components of Hot Dark Matter (HDM) and atmospheric neutrino
oscillations seems to prefer a scenario in which two or three light
neutrinos are nearly degenerate and have mass in the O(eV) range.
Recent data on neutrinoless double beta decay (or absence
thereof) \cite{2beta} appear
to rule out Majorana neutrinos heavier than 0.2 eV, at least in
the simplest versions.
Here it will be shown how, in our scenario, one can obtain
three near-degenerate neutrinos whose mass can be of the order
of a few eV's and is of the Dirac type. Consequently, in our
model, there will be {\em no} neutrinoless double beta decay,
and hence no contraint on the Dirac neutrino masses from such
a search.

As we have discussed above, Option 1 contains no family-singlet
fermion field and freedom from the Witten anomaly dictates
that the number of families should be even. Furthermore, we have argued 
that this even number should be four. As a result, the gauge group
for this option is:
\begin{equation}
SU(3)_c \otimes SU(2)_L \otimes U(1)_Y \otimes SO(4) \otimes SU(2)_{\nu_R}
\label{gaugegroup}
\end{equation}
The reader is referred to Table 1 for a list of particles that participate
in this model.

\subsection{Computation of the diagonal elements of the $4 \times 4$
neutrino mass matrix}
\label{diagonal}

Without the extra vector-like fermions, $F$, $M_1$ and $M_2$, the only
gauge-invariant Yukawa coupling involving leptons would be
${\cal L}_Y = g_E \bar{l}_L^{\alpha} \phi e_{\alpha\, R} + H.c.$,
(where $\alpha= 1,..,4$ is the family index), giving rise to
equal masses for the charged leptons. 
Unbroken $SU(2)_{\nu R}$
forbids a similar term for the neutrinos and they remain massless
at this level. (Notice that, since
we are only interested in Dirac neutrino masses, a gauge-invariant
Majorana mass term of the type $\eta^{i\,\alpha}_{R} \eta_{i\,\alpha\,R}$
is forbidden by $L$ symmetry.)
We {\em know} that the charged
leptons are not degenerate in mass. We also {\em know} that the width 
of the Z boson \cite{LEP}
constrains the mass of the fourth neutrino to be 
larger than half the Z mass. This is where the vector-like fermions
listed in Table 1 come in. Because of their vector-like nature,
they can have {\em arbitrary} gauge-invariant
bare masses. It is seen below that some of these masses can be as
low as a few hundreds GeVs and are thus accessible to 
future experimental searches.
 
The Yukawa part of the Lagrangian involving leptons can be
written as
\begin{eqnarray}
{\cal L}^Y_{Lepton}& =& g_E \bar{l}_L^{\alpha} \phi e_{\alpha\, R} +
G_1 \bar{l}^{\alpha}_{L} \Omega_{\alpha} F_{R} +
G_{M_1} \bar{F}_{L} \phi {\cal M}_{1R}+
G_{M_2} \bar{F}_{L} \tilde{\phi} {\cal M}_{2R} +
G_2 \bar{{\cal M}}_{1L} \Omega_{\alpha} e^{\alpha}_{R} + \nonumber \\
          &  &G_3 \bar{\cal{M}}_{2L} \rho^{\alpha}_{m} \eta^{m}_{\alpha R}
+ M_F \bar{F}_L F_R + M_1 \bar{{\cal M}}_{1L} {\cal M}_{1R} +
M_2 \bar{{\cal M}}_{2L} {\cal M}_{2R} + h.c.
\label{lag1}
\end{eqnarray}
The assumption of an unbroken $L$ symmetry
forbids the presence of Majorana mass terms as mentioned above.
 
Notice that the values of $M_{F,1,2}$ are arbitrary.
What they might be will be the subject of the discussion
presented below.
After integrating out the $F$, ${\cal M}_1$, and ${\cal M}_2$ fields,
the relevant part of the effective Lagrangian below $M_{F,1,2}$ reads
\begin{eqnarray}
{\cal L}^{Y,eff}_{Lepton}& =& g_E \bar{l}_L^{\alpha} \phi e_{\alpha \,R} +
G_E \bar{l}^{\alpha}_{L} (\Omega_{\alpha} \phi \Omega^{\beta}) e_{\beta \,R} +\nonumber \\
          &  &G_N \bar{l}^{\alpha}_{L}(\Omega_{\alpha} \tilde{\phi} \rho^{\beta}_{i}) 
\eta^{i}_{\beta \,R} + H.c. ,
\label{lag2}
\end{eqnarray}
where
\begin{equation}
G_E =\frac{G_1 G_{M_1} G_2}{M_F M_1};\, G_N =\frac{G_1 G_{M_2} G_3}{M_F M_2}.
\label{Yuka1}
\end{equation} 
This is a tree-level effective Lagrangian whose consequences are now
presented.

Let us discuss the implication of each term on the right-hand side of Eq. (\ref{lag2}).
As stated in the preceding paragraph, the first term gives rise to
equal masses for the charged leptons. The second term would lift the
degeneracy of the charged lepton sector once $\Omega$ acquires a vacuum
expectation value (VEV). The third term gives rise to a neutrino mass
once {\em both} $\Omega$ and $\rho$ acquire a VEV. It is clear that, in
our model, neutrino masses can appear only when {\em both} $SO(4)$ and
$SU(2)_{\nu_R}$ are spontaneously broken while the charged lepton masses
are non zero (but equal) even if $SO(4)$ is unbroken. Only when $SO(4)$
is broken will the charged lepton mass degeneracy be lifted. 


Let us assume: $<\Omega> = (0,0,0,V)$ and $<\rho> = (0,0,0,V^{\prime} \otimes s_1)$,
where $s_1 = \left( \begin{array}{c} 1 \\ 0 \end{array} \right)$. 
Notice that each component (under
$SO(4)$) of $\rho$ transforms as a doublet under $SU(2)_{\nu_R}$. If we denote
the 4th element of $\eta_R$ by $(N_R,\,\tilde{N}_{R})$, one can use
the above two VEV's along with $<\phi> = (0,\,v/\sqrt{2})$ ( $v \approx$ 246 GeV) 
in Eq.(\ref{lag2}) to write down a Dirac mass term for the 4th generation neutrino, namely
\begin{equation}
\tilde{G}_N \frac{v}{\sqrt{2}} \bar{N}_L N_R + h.c.;\,
\tilde{G}_N = G_1 G_{M_2} G_3 \frac{V\,V^{\prime}}{M_F\,M_2},
\label{G_N}
\end{equation}
giving
\begin{equation}
m_N = \tilde{G_N} \frac{v}{\sqrt{2}}.
\label{m_N}
\end{equation}
At {\em tree level}, all other neutrinos are massless. Their
masses arise at the one-loop level as shown below. The
{\em Dirac} mass of the fourth neutrino
could be rather {\em heavy}. In fact, it is not
unreasonable to expect $G_1$, $G_{M_2}$ and $G_3$ to be of the order of unity.
In consequence, as long as
\begin{equation}
V\, V^{\prime}/M_F\,M_2 \sim O(1), 
\end{equation}
one might expect the fourth neutrino to
be even as heavy as 175 GeV. Certainly, the LEP bound of $M_Z/2$ can easily
be satisfied.

Why are the other three neutrinos massless at tree level? Firstly, it
is so because, from Eq. (\ref{lag1}) and Eq. (\ref{lag2}), one 
can see that, after integrating
out the heavy vector-like fermions, there is no (tree-level, dimension 6) 
operator which contains,as a factor, a term such 
as $\bar{l}^{m}_L \tilde{\phi} \eta^{m}_{iR}$,
where $m=1,2,3$ is a family index, which would give rise to a mass term
for the three light neutrinos. An effective (dimension 6) operator
which contains the aforementioned term would necessarily come from
a loop integration such as the one shown in Fig. 1. Just like the various
terms which appear in Eq. (\ref{lag2}), this effective operator would also
contain the scalar fields $\Omega$ and $\rho$. It would appear as
\begin{equation}
\bar{l}^{m}_L \tilde{\phi} \eta^{m}_{iR} (\Omega_{\alpha} \rho^{\alpha\,i}).
\end{equation}
As pointed out in the Appendix, a term such as
$(\Omega^{\alpha} \rho_{\alpha\,i})$ appears as part of a quartic term in 
the potential which explicitely breaks the extra global symmetry that
the scalar sector posesses. As a result, the extra NG bosons are, in
fact, pseudo NG bosons and acquire a mass which is proportional
to the coupling $\lambda_4$ as shown in Eq. (\ref{Omegarhomass}) 
of the Appendix. 


In order to compute the one-loop contributions to neutrino
masses, let us recall, in this section, the results obtained
in the Appendix concerning the relevant mass eigenstates in 
the scalar sector. We have
\begin{mathletters}
\label{physscal}
\begin{equation}
H_4 = \cos \alpha \tilde{H}_4 - \sin \alpha \tilde{h}_4,
\end{equation}
\begin{equation}
h_4 = \sin \alpha \tilde{H}_4 + \cos \alpha \tilde{h}_4,
\end{equation}
\begin{equation}
\Omega_i = \cos \beta \tilde{\Omega}_i - \sin \beta Re\tilde{\rho}_i,
\end{equation}
\begin{equation}
Re \rho_i = \sin \beta \tilde{\Omega}_i + \cos \beta Re\tilde{\rho}_i,
\end{equation}
\end{mathletters}
where $i=1,2,3$ and where the states with the $\tilde{}$ sign are
mass eigenstates. The Yukawa couplings which will be involved
in the computation of neutrino masses can now be written in terms
of the mass eigenstates. For example, $G_1 \bar{l}^{\alpha}_{L} 
\Omega_{\alpha} F_{R}$ can be written as
\begin{equation}
G_1 \bar{l}^{4}_{L} \Omega_{4} F_{R} = 
G_1 \bar{l}^{4}_{L} (\cos \alpha \tilde{H}_4 - \sin \alpha 
\tilde{h}_4) F_{R},
\label{omega4}
\end{equation}
\begin{equation}
G_1 \bar{l}^{i}_{L} \Omega_{i} F_{R} = 
G_1 \bar{l}^{i}_{L} (\cos \beta \tilde{\Omega}_i - \sin 
\beta Re\tilde{\rho}_i) F_{R}.
\label{omegai}
\end{equation}
where $i=1,2,3$.
Also, $G_3 \bar{M}_{2L} \rho^{\alpha}_{m} \eta^{m}_{\alpha R}$ ($m=1,2$)
can be now written as
\begin{equation}
G_3 \bar{M}_{2L} \rho^{4}_{1} \eta^{1}_{4 R}=
G_3 \bar{M}_{2L} (\sin \alpha \tilde{H}_4 + 
\cos \alpha \tilde{h}_4 + i Im \rho_4)_{1}\eta^{1}_{4 R},
\label{yuka2}
\end{equation}
\begin{equation}
G_3 \bar{M}_{2L} \rho^{i}_{1} \eta^{1}_{i R}=
G_3 \bar{M}_{2L}(\sin \beta \tilde{\Omega}_i + 
\cos \beta Re\tilde{\rho}_i + i Im \rho_i)_{1} \eta^{1,i}_{R}.
\label{yuka3}
\end{equation}
The above equations, in addition to $G_{M_2} \bar{F}_{L} \tilde{\phi} M_{2R}$,
form the basis for constructing the one-loop diagrams as shown
in Fig.1. As one can immediately see, the only scalars that
participate in the loop integration are $\tilde{H}_4$,
$\tilde{h}_4$, $\tilde{\Omega}_i$, and $\tilde{\rho}_i$.  
The contributions to the light neutrino
masses will contain a factor $\cos \beta \sin \beta
= \sin(2\beta)/2$ for 
$\tilde{\Omega}_i$ and -$\cos \beta \sin \beta$ for $Re \tilde{\rho}_i$.

The masses of the physical Higgs scalars, $H_4$ and $h_4$, and
those of the pseudo NG bosons, $Re \tilde{\rho}_i$ ($i=1,2,3$),
are given by Eqs. (\ref{physhiggs},\ref{Omegarhomass}) in the 
Appendix. Since the one-loop contributions to
the 4th neutrino mass are expected to be small compared with its
tree-level value, we shall concentrate in this section on the
light neutrino masses. There we shall be concerned only with
$\tilde{\Omega}_i$ (NG bosons) and $Re \tilde{\rho}_i$ (pseudo NG bosons) 
($i=1,2,3$). In the 'tHooft-Feynman gauge, the NG bosons will have
a propagator with a mass which is that of the family gauge bosons. We
shall denote it by $M_G$. We shall call the mass of the pseudo NG bosons,
$M_P$. 


The result obtained from the diagrams as shown in Fig. 1 for the three
light neutrinos is
\begin{equation}
m_\nu = \tilde{G}_{\nu} \frac{v}{\sqrt{2}},
\label{mnu}
\end{equation}
where
\begin{equation}
\tilde{G}_{\nu} = G_1 G_{M2} G_3 \frac{\sin(2\beta)}
{32\,\pi^2}\,(I(\tilde{\Omega})- I(Re\tilde{\rho})),
\label{gnu}
\end{equation}
and where
\begin{equation}
I(\tilde{\Omega})- I(Re\tilde{\rho}) = \frac{1}{M_F-M_2}
\{\frac{M_F[M_F^2(M_G^2\ln(\frac{M_G^2}{M_F^2})-M_P^2\ln
(\frac{M_P^2}{M_F^2}))+ 
M_G^2 M_P^2 \ln(\frac{M_P^2}{M_G^2})]}{(M_G^2-M_F^2)(M_P^2-M_F^2)} -(M_F 
\leftrightarrow M_2)\}.
\label{int1}
\end{equation}
For notational convenience, we shall define:
\begin{equation}
\Delta I(G,P) \equiv I(\tilde{\Omega})- I(Re\tilde{\rho}),
\label{DeltaGP}
\end{equation}
It is convenient to express the mass of the light
neutrinos by the following ratio:
\begin{equation}
\frac{m_\nu}{m_N} = \frac{M_F M_2}{V V^{\prime}}\frac{\sin(2\beta)}
{32\,\pi^2}\,\Delta I(G,P),
\label{rat1}
\end{equation}
where $m_N$ is defined by Eq.\ (\ref{m_N}).

One should mention for completeness the tiny one-loop contribution
to the 4th neutrino mass. If we denote by this contribution by 
$\delta m_4$, it is straigthforward to see that it is
given precisely by the same formula for the light neutrino mass,
Eq. (\ref{mnu}), with the following replacements: $\beta \rightarrow
\alpha$, $M_G \rightarrow M_{H_4}$, $M_P \rightarrow
M_{h_4}$, namely
\begin{mathletters}
\label{delm4}
\begin{equation}
\delta m_4 = \tilde{G}_4 \frac{v}{\sqrt{2}},
\end{equation}
\begin{equation}
\tilde{G}_4 = G_1 G_{M2} G_3 \frac{\sin(2\alpha)}
{32\,\pi^2}\,\Delta I(G,P),
\end{equation}
\end{mathletters}
where the form of $I(\tilde{H}_4)- I(\tilde{h}_4)$ is
identical to Eq. (\ref{int1}) with the replacements as mentioned
above. This contribution will play an insignificant role
in the mass matrix, but it has to be mentioned for
completeness.

The above results were obtained at one loop. One wonders
if higher loop contributions might be significant. It turns
out that, because of the nature of the interactions, the
next correction occurs at the three loop level. It means
that the correction to the one-loop light neutrino mass
is at the two-loop order. Considering that already the
one-loop result is O($<10^{-10}$), a two-loop correction
to that result would most likely be insignificant, even
for the mass splitting to be discussed below. Above all,
the experimental results are far from being precise enough
to even contemplate such a tiny correction. From hereon, we 
shall assume that these three-loop corrections are
insignificant in the computation of the mass splittings.

At this stage, the three light neutrinos are degenerate. A
discussion of the lifting of the degeneracy will follow a
more general discussion of the implications of Eq. (\ref{rat1}). It is
clear that the ``light family'' symmetry would have to be
broken in order for the ``light'' fermions to mix. It is
also clear that the neutrino masses (one heavy and three light)
derived so far represent only the diagonal elements of a
$4 \times 4$ neutrino mass matrix. If the discussion 
presented in this section on light neutrino masses is
to be at all interesting, it is imperative to assume that
the bulk of at least one, if not all, of the light neutrino 
masses comes from Eq. (\ref{mnu}). 

At this point, an important remark is in order here. As we
have stressed above, the near-degeneracy of the light
neutrinos in no way implies that a similar situation will
occur in the charged lepton sector. In fact, we will show
in a separate paper that this will not be the case.

Under what conditions will $\tilde{G}_{\nu}$ be of the order of
$10^{-11}$ or less? First of all, as we have seen from Eq. (\ref{m_N}),
in order to have a ``heavy'' fourth neutrino, one should have
$G_1 G_{M2} G_3 \frac{V\,V^{\prime}}{M_F\,M_2}\approx O(1)$. 
This puts a condition on the
angle $\beta$ itself, namely ($\tan \beta \equiv V^{\prime}/V$)
\begin{equation}
\tan \beta \approx \frac{1}{G_1 G_{M2} G_3} \frac{M_F M_2}{V^2}.
\label{beta1}
\end{equation}
As we have stated earlier, it is not unreasonable to assume that 
$G_1$, $G_{M_2}$ and $G_3$ to be of the order of unity.
With $M_G^2 \sim g^2 V^2$ (where $g$ is the $SO(4)$ gauge coupling), 
Eq. (\ref{beta1}) becomes
\begin{equation}
\tan \beta \approx g^2 \frac{M_F}{M_G}\frac{M_2}{M_G}.
\label{beta2}
\end{equation}
The above estimate for the constraint on the angle $\beta$ will be used
in our computation of the light neutrino masses. With this in mind,
we can now proceed to make an estimate of the ratio $m_\nu/m_N$, where now
$V\, V^{\prime}/M_F\,M_2 \sim O(1)$ and Eq. (\ref{rat1}) becomes
\begin{equation}
\frac{m_\nu}{m_N} = \frac{\sin(2\beta)}
{32\,\pi^2}\,(I(\tilde{\Omega})- I(Re\tilde{\rho})).
\label{rat2}
\end{equation}
As we have seen above, the result (\ref{rat2}) depends only on 
ratios of masses of the particles in the loop integral and
not on their absolute values. Because of that fact, the results will be
shown in units of $M_F$ which can be as small or as large as one wishes.

Before moving on to discuss the implications of Eqs. (\ref{rat1}) 
and (\ref{rat2}), one remark
is in order here. From Eq. (\ref{int1}), one can see that the light neutrino mass
vanishes when $M_G = M_P$. Since there is no reason (as far as the
present construction of the model is concerned) for this equality
to be valid, we shall dismiss this possibility. We shall concentrate
instead on the criteria for having small $m_\nu$ for arbitrary
$M_G$ and $M_P$ (and $M_F$ and $M_2$ as well).

The results are shown in Figs. 2, 3, 4 and 5. A few comments are in
order here. First of all, as we have mentioned above, our results depend
on ratios of the four masses which enter the loop integral: $M_F$,
$M_P$, $M_G$, and $M_2$. One can symbolically denote one of the masses
as $M=1$, and the other three will be multiples of that chosen one.
Which one should be chosen is a matter of convenience and phenomenological
interest. In particular, we choose $M_F=1$ because there is a possibility
that the vector-like fermions $F$  could be detected if their masses are low
enough. 

A glance at Figs. 2-5 reveal that it is relatively easy to obtain
a very small ratio $R \equiv m_\nu/m_N$. In particular, one can see that
large values of $M_2$, the mass of the singlet fermion field ${\cal M}_2$, 
are sufficient
to obtain small values for $R \equiv m_\nu/m_N$. For instance, one can see 
that, roughly speaking, $R \equiv m_\nu/m_N \lesssim 10^{-11}$ when
$M_2 \gtrsim 10^6$ (in units of $M_F$). Although conceptually quite different,
the above fact is very reminescent of the see-saw mechanism in that there
is one large scale: Majorana for see-saw, $M_2$ for this scenario, and
one ``small'' scale: Dirac mass $m_D$ for see-saw, $M_F$ for this scenario.
The important point that we wish to make is the fact that the general
result obtained here, namely the smallness of light neutrino masses, does
not depend on one particular combination of masses which would imply fine tuning,
a point which was not made quite clear in Ref. \cite{hung}, 
but only on ``large'' ratio
of masses whatever they might be. In this sense, the smallness of neutrino
masses in our scenario is no less natural than the ones obtained from the
see-saw mechanism.

In Figs. 2-5, we show the results for the case $M_2 > M_G$. There
is, of course, absolutely no reason for this ordering. It is a matter of
presentation. We obtain exactly the same results with the roles of $M_2$ and
$M_G$ reversed. As can be inferred from the figures, for a given value of $M_P$
($M_F =1$), $R \equiv m_\nu/m_N \lesssim 10^{-11}$ if the ratio
$M_G/M_2$ is below a certain value. For example, for $M_G \lesssim 10^5$,
one has $M_G/M_2 \lesssim 10^{-3}$, while for $M_G \gtrsim 10^7$, one
has $M_G/M_2 \approx 10^{-2} - 10^{-1}$. What this says is that the larger
the mass is (e.g. $M_G$), the less mass splitting is needed in order to have a
small $R$. 

At this stage, we can only say that $m_\nu$ can be very small. What we
cannot say is exactly what its value should be. This should come from
some deeper theory. Instead, we shall use present constraints to
restrict the range of values for $M_{G,P,2}$.

Having seen how one can obtain {\em very} small $m_\nu$, the next question
would be: How small can one allow $m_\nu$ to be if one takes into account
the neutrino oscillation data? First of all, atmospheric neutrino
oscillation data gives a difference of mass squared $\Delta m^2
\approx 10^{-3} eV^2$ while solar neutrino oscillation data gave
$\Delta m^2 \approx 10^{-5} eV^2$ (MSW) or $10^{-10} eV^2$ (vacuum).
In anticipation of new data, the LSND results are not taken into
account in our rough estimation of various mass scales.
Without any need for a specific model, one can say that the atmospheric
data implies that at least one of the three neutrinos should have a mass
of {\em at least} $3 \times 10^{-2} eV$, while the solar data implies
that at least one of the remaining two should have a mass of
{\em at least} $3 \times 10^{-3} eV$ (MSW) or $10^{-5} eV$ (vacuum).
As we have seen above, the 4th neutrino can be quite heavy. For the
sake of argument, let us assume here that its mass is approximately
100 GeV. Since our three light neutrinos are practically degenerate
-a lifting of which will be discussed below, the atmospheric data
alone constrains $R$ to be greater than  approximately $10^{-14}$.
This in turn constrains $M_2 \lesssim 10^{12}$ (in units of $M_F$)
for the case $M_2>M_G$, or $M_G \lesssim 10^{12}$ for the reverse case.
Notice that this rough estimate is only for illustration purpose. 

There is however one interesting
piece of information which could be quite interesting, phenomenologically
speaking: the presence of vector-like fermions which carry weak quatum
numbers and which could be relatively ``light''. These are the fermions
$F$ with mass $M_F$ as indicated above. Let us recall from the above
discussions that $M_{G,P,2}$ are all expressed in units of $M_F$ which
itself could take on any value, even a few hundreds of GeV. The sole 
restriction will be from experimental constraints, a subject to which we 
shall come back below. Furthermore, we can see from the results that
the mass of the pseudo-NG bosons can also be ``low'' as well (Fig.1)
which could provide a further experimental clue.

We now turn to an important issue: the lifting of the mass degeneracy
of the light neutrinos. The analysis presented below will reveal quite
interesting implications such as the correlation between the actual values
of the masses and $\Delta m^2$, which can have a profound cosmological
consequence. For neutrino masses which are large enough to provide
part of HDM, the MSW solution of the solar neutrino problem is preferred.
If the vacuum solution turns out to be the correct one, the neutrino masses
will be much too light in our scenario to play a role in HDM.

We shall divide the discussion presented below into two parts.
First we analyze the case when there is no mixing between the 4th
neutrino and the lighter three. It will be seen that an interesting
feature emerges: $\Delta m_{23}^2 \approx \Delta m_{21}^2$-a quasi-symmetric
splitting. ($\Delta m_{31}^2$ is of the same order.) This phenomenon
could be called a mass splitting quasi-degeneracy. Of course, solar and
atmospheric neutrino data suggest otherwise. Next, we will show
how this mass splitting quasi-degeneracy can be lifted, suggesting-
at least in our scenario-the presence of a 4th neutrino.

In what follows, we will neglect any possible CP phase in the
neutrino mass matrix since we will be concerned only with
$\Delta m^2$ and present data on neutrino oscillations are not
sensitive to the presence of such a phase. In addition, we
shall concentrate in the next two subsections only on
$\Delta m^2$. A full comparison with the data will necessitate
the inclusion of the leptonic ``CKM'' angles coming
from $V_L = U_{l}^{\dagger}U_{\nu}$. In the two subsections
presented below, we shall see what $U_{\nu}$ might look like.
To complete the discussion, we shall use a model for
$U_{l}$ in order to make some statements about the size of
the mixing angles. The subject of the charged lepton mass matrix
itself will be dealt with in a subsequent publication.

\subsection{Neutrino mass matrix I: What if there is no mixing
between the 4th and the lighter three neutrinos? }
\label{matrix1}
 
The $4 \times 4$ neutrino mass matrix obtained at this point is purely
{\em diagonal}. We would like to examine how mass mixing might arise.
In particular, we would like to lift the degeneracy of the three light
neutrinos. In this section we will concentrate on the scenario where
there is mass mixing only among the three light neutrinos. We will show
that, in this scenario, $\Delta m_{23}^2 \approx \Delta m_{21}^2$.
If this were experimentally the case, it would be hard to detect the
influence of the 4th neutrino  since it does not mix with the other three.
Since the atmostpheric and solar data appear to point to 
$\Delta m_{23}^2 \gg \Delta m_{21}^2$, we will present in the next
section what can be done in order to be in agreement with the data.
It turns out that this can be accomplished if one introduces a mixing
with the 4th neutrino. This implies that, at least in our model, 
$\Delta m_{23}^2 \gg \Delta m_{21}^2$ implies the existence of a
4th neutrino, and hence a 4th generation.

The degeneracy of the three light neutrinos at this level comes from the
fact that there is a remaining global $SO(3)$ symmetry which manifests itself 
through the equality of the masses of the family gauge bosons ($M_G$) as well as
those of the pseudo-NG bosons ($M_P$). It is then clear that one needs to
break that remaining global symmetry in order to remove the degeneracy of
the light neutrino masses. We would want to do this in such a way as to
preserve the quasi-degeneracy of the light neutrinos. There are probably several
ways to achieve this, and we will present one of them here.

Since we have seen how the diagonal elements of the neutrino mass matrix for the
three light neutrinos are obtained at the one loop level, it is natural to
envision a scenario in which the mixings themselves are obtained at {\em one loop}.
A look at Figs. 1 reveals that the most ``straightforward'' way to induce
mixings at one loop is for $\tilde{\Omega}_i$ and/or $Re \tilde{\rho}_i$
to have mixed couplings, i.e. to both $\nu_{Li}$ and $\nu_{Lj}$ as well as
to both $\eta_{Ri}$ and $\eta_{Rj}$. This could come from mixings among
$\tilde{\Omega}_i$ with different family indices and/or the mixings among
$Re \tilde{\rho}_i$. Before getting into the details of what kinds of
interactions are needed to break the remaining global $SO(3)$ symmetry
and hence inducing the mixings, it is instructional to assume that
such a mixing among the boson masses occurs and to write down the Yukawa
couplings (\ref{yuka2},\ref{yuka3}) in terms of the new boson mass eigenstates.

Let us first look at the states $\tilde{\Omega}_i$. As we have discussed
earlier, these are the NG bosons which are absorbed by the corresponding
family gauge bosons. When these NG bosons get mixed,
there will be mass mixings among the corresponding family gauge bosons.
Let us denote the orthogonal matrix which diagonalizes these family gauge bosons by
$A_{\Omega}$. We shall choose the following representation for $A_{\Omega}$:
\begin{equation}
A_{\Omega} = \left(
\begin{array}{ccc}
c_2 c_3 & -s_1 s_2 c_3 + c_1 s_3 & c_1 s_2 c_3 + s_1 s_3 \\
-c_2 s_3 & c_1 c_3 + s_1 s_2 s_3 & -c_1 s_2 s_3 + s_1 c_2 \\
-s_2 & -s_1 c_2 & c_1 c_2
\end{array}
\right)\
\label{Aomega}
\end{equation}
where $c$ and $s$ represent the cosine and sine. If we denote by
$\tilde{\Omega}_{i}^{\prime}$ the logitudinal components of the gauge
boson mass eigenstates, its relationship with $\tilde{\Omega}_i$ in the
unmixed case is given by
\begin{equation}
\left(
\begin{array}{c}
\tilde{\Omega}_{1}\\
\tilde{\Omega}_{2}\\
\tilde{\Omega}_{3}
\end{array}
\right) = A_{\Omega}^{T} \left(
\begin{array}{c}
\tilde{\Omega}_{1}^{\prime}\\
\tilde{\Omega}_{2}^{\prime}\\
\tilde{\Omega}_{3}^{\prime}
\end{array}
\right)\
\label{omega}
\end{equation}
where $A_{\Omega}^{T}$ is given by
\begin{equation}
A_{\Omega}^{T} = \left(
\begin{array}{ccc}
c_2 c_3 &-c_2 s_3 & -s_2\\ 
-s_1 s_2 c_3+c_1 s_3 & c_1 c_3 + s_1 s_2 s_3 & -s_1 c_2\\
c_1 s_2 c_3 + s_1 s_3  & -c_1 s_2 s_3 + s_1 c_2 & c_1 c_2
\end{array}
\right)\
\label{AT}
\end{equation}
The masses of the corresponding gauge bosons are now denoted by
\begin{equation}
M_{G_1}^2 = M_G^2 + \delta_1;\, M_{G_2}^2 = M_G^2 + \delta_2;\,
M_{G_3}^2 = M_G^2 ,
\label{Gmass}
\end{equation}
where $\delta_{1,2}$ can be positive
or negative. Notice that $\delta_{1,2}$ and the mixing angles shown
above are {\em related}, i.e. they are all derived from the same
boson mass matrix. We will show an example of such fact below.

We can now replace the unprimed states in Eqs.(\ref{omegai},\ref{yuka3}) 
by the primed states using
Eq. (\ref{omega}). We can then compute the one-loop contributions to the elements
of the neutrino mass matrix ${\cal M}_{\nu}$. Let us first look at
the contributions to the light neutrino masses and mixings coming
from the $\tilde{\Omega}_i$ states. The two terms which are crucial
for this computation are
\begin{equation}
G_1 \bar{l}^{i}_{L} \cos \beta \tilde{\Omega}_{i} F_R =
G_1 \bar{l}^{i}_{L} \cos \beta A_{\Omega,i}^{T,j} 
\tilde{\Omega}^{\prime}_{j} F_R
\end{equation}
and
\begin{equation}
G_3 \bar{M}_{2L}\sin \beta \tilde{\Omega}_{i} \eta^{1,i}_{R} =
G_3 \bar{M}_{2L}\sin \beta \tilde{\Omega}^{\prime}_{j}
A_{\Omega,i}^{j} \eta^{1,i}_{R}
\label{yuka4}
\end{equation} 
In the loop integrations, one will encounter the following propagators:
\begin{mathletters}
\label{prop}
\begin{equation}
\frac{1}{k^2- M_{G3}^2} = \frac{1}{k^2 - M_G^2},
\end{equation}
\begin{equation}
\frac{1}{k^2 - M_{G1}^2} = \frac{1}{k^2 - M_G^2} + \frac{\delta_1}
{(k^2- M_{G2}^2)(k^2 - M_G^2)} ,
\end{equation}
\begin{equation}
\frac{1}{k^2 - M_{G2}^2} = \frac{1}{k^2 - M_G^2} + \frac{\delta_2}
{(k^2- M_{G2}^2)(k^2 - M_G^2)} ,
\end{equation}
\end{mathletters}

With the above remarks in mind, let us proceed to calculate the contributions
of $\tilde{\Omega}^{\prime}$ to the neutrino mass matrix. We shall concentrate 
first on the $3\times3$ submatrix of the light neutrino sector. As a prelude
to the computation of the full submatrix, let us show how two elements are
calculated: ${\cal M}_{\nu}^{11}$ and ${\cal M}_{\nu}^{12}$. In these
computations. we shall use, as an example, the explicit form for $A_{\Omega}$
shown in Eq. (\ref{Aomega}). For the complete calculations of the matrix elements,
we shall use the notations $A_{ij}$ for $A_{\Omega}$.

a) In the calculation of the contribution of $\tilde{\Omega}^{\prime}$ to
${\cal M}_{\nu}^{11}$, one combines Eq. (\ref{omega}) with Eq. (\ref{AT}) to 
get the following combination of $\tilde{\Omega}^{\prime}$:
\begin{equation}
(c_2 c_3 \tilde{\Omega}_{1}^{\prime} -c_2 s_3 \tilde{\Omega}_{2}^{\prime} 
-s_2 \tilde{\Omega}_{3}^{\prime})^2,
\end{equation}
which gives the following combination of propagators:
\begin{equation}
c_{2}^{2} c_{3}^{2} \langle \tilde{\Omega}_{1}^{\prime}\tilde{\Omega}_{1}^{\prime}
\rangle + c_{2}^{2} s_{3}^{2}
\langle \tilde{\Omega}_{2}^{\prime}\tilde{\Omega}_{2}^{\prime}\rangle +
s_{2}^2 \langle \tilde{\Omega}_{3}^{\prime}\tilde{\Omega}_{3}^{\prime}\rangle
\label{angle}
\end{equation}
Upon using the propagators listed in Eqs.(\ref{prop}) in the one-loop integral (Fig.1),
one obtains the following replacement (the reader is referred to Eq. (\ref{gnu}) for
a comparison):
\begin{equation}
\frac{\sin(2\beta)} {32\,\pi^2}\,I(\tilde{\Omega}) \rightarrow
\frac{\sin(2\beta)} {32\,\pi^2}(I(\tilde{\Omega}) + c_2^2 c_3^2 \delta_1
I(M_G,M_{G1}) + c_2^2 s_3^2 \delta_2 I(M_G,M_{G2}))
\label{replace}
\end{equation}
where ($i=1,2$)
\begin{equation}
\delta_i I(M_G,M_{Gi})=  \frac{1}{M_F-M_2}
\{\frac{M_F[M_F^2(M_G^2\ln(\frac{M_G^2}{M_F^2})-M_{Gi}^2\ln
(\frac{M_{Gi}^2}{M_F^2}))+ 
M_G^2 M_{Gi}^2 \ln(\frac{M_{Gi}^2}{M_G^2})]}{(M_G^2-M_F^2)(M_{Gi}^2-M_F^2)} -(M_F 
\leftrightarrow M_2)\}.
\label{deltaI2}
\end{equation}
One can see that, in the symmetry limit where $\delta_i \rightarrow 0$
($M_{Gi} \rightarrow M_G$), $\delta_i I(M_G,M_{Gi})$ vanishes identically.

One interesting remark worth mentioning is the following: In (\ref{replace}), the
first term $I(\tilde{\Omega})$ contains no mixing angles. In fact,
the coefficient in front of $I(\tilde{\Omega})$ is $c_2^2 c_3^2 +
c_2^2 s_3^2 + s_2^2 = 1$, which is the result of $A_{\Omega}$
being an orthogonal matrix.

We do not give the explicit form for $I(\tilde{\Omega})$ because, after
taking into account the contribution of $Re\tilde{\rho}_i$, one
obtains the combination $I(\tilde{\Omega}) - I(Re\tilde{\rho})$ which
is already given by Eq. (\ref{int1}).

When the boson mass differences, represented by $\delta_i$, are small
compared with $M_G^2$, another useful form which could be used is
given by ($i=1,2$)
\begin{equation}
\delta_i I(M_G,M_{Gi})=  -x_i I(M_G, x_i)
\label{deltaI3}
\end{equation}
where
\begin{equation}
I(M_G, x_i) = \frac{M_G^2}{M_F-M_2}
\{\frac{M_F[-M_F^2 (1 + x_i + \ln(\frac{M_G^2}{M_F^2})) + M_G^2 (1+ x_i)]}
{(M_G^2-M_F^2)^{2}(1+x_{i}(M_G^2/(M_G^2-M_F^2)))} -(M_F 
\leftrightarrow M_2)\},
\label{int2}
\end{equation}
and where
\begin{equation}
x_i = \frac{\delta_i}{M_G^2},
\end{equation}
so that
\begin{equation}
M_{G3}^2 = M_G^2;\, M_{G1}^2 = M_G^2 (1 + x_1);\,  M_{G2}^2 = M_G^2 (1 + x_2).
\label{Gmass2}
\end{equation}
Here one could explicitely see the vanishing of  $\delta_i I(M_G,M_{Gi})$
in the symmetry limit because of the explicit appearance of $\delta_i$
on the right-hand side of the equation.

The other diagonal elements of the neutrino mass matrix can be analogously
calculated. One just needs to replace the combination of angles in (\ref{angle})
with the appropriate ones.

b) For the 1-2 element, the appropriate combination of propagators is given
by
\begin{equation}
c_{2} c_{3}( -s_1 s_2 c_3 + c_1 s_3) \langle \tilde{\Omega}_{1}^{\prime}
\tilde{\Omega}_{1}^{\prime}\rangle - c_2 s_3(c_1 c_3 + s_1 s_2 s_3)
\langle \tilde{\Omega}_{2}^{\prime}\tilde{\Omega}_{2}^{\prime}\rangle +
s_1 s_2 c_2 \langle \tilde{\Omega}_{3}^{\prime}\tilde{\Omega}_{3}^{\prime}\rangle
\end{equation}
It is now straigthforward to compute ${\cal M}_{\nu}^{12}$. It is given by
\begin{equation}
{\cal M}_{\nu}^{12}(\tilde{\Omega}) =\frac{\sin(2\beta)} 
{32\,\pi^2}(c_{2} c_{3}( -s_1 s_2 c_3 + c_1 s_3) \delta_1 I(M_G,M_{G1})
- c_2 s_3(c_1 c_3 + 
s_1 s_2 s_3)\delta_2 I(M_G,M_{G2})),
\label{m12}
\end{equation}
where we have the appearance of the same $\delta_i I(M_G,M_{Gi})$.
Notice that ${\cal M}_{\nu}^{12}(\tilde{\Omega})$ denotes the contribution
coming from $\tilde{\Omega}$ only. The full element will also include
the contribution coming from $Re\tilde{\rho}$.

Notice that the term $I(\tilde{\Omega})$ is not present in (\ref{m12}). 
Again this
is due to the orthogonality of $A_{\Omega}$. The coefficient appearing
in front of $I(\tilde{\Omega})$ is $c_{2} c_{3}( -s_1 s_2 c_3 + c_1 s_3)
- c_2 s_3(c_1 c_3 + s_1 s_2 s_3)+ s_1 s_2 c_2 =0$. The orthogonality of
$A_{\Omega}$ implies that the product of any two columns is equal to zero.
As a result we can see that, in the symmetry limit, ${\cal M}_{\nu}^{12}$
vanishes identically. This applies to all the other off-diagonal elements.

In order to complete the computation of the matrix elements (including
the 1-1 and 1-2 elements), one has to
say something about the contributions coming from the pseudo-NG bosons themselves.
One might imagine that the same mechanism which breaks the global
$SO(3)$ symmetry also induces mixing among the degenerate pseudo-NG bosons.
We will {\em assume} that the same matrix $A_{\Omega}$ diagonalizes the
pseudo-NG boson sector so that, instead of the combination of $\tilde{\Omega}_i$ and 
$Re\tilde{\rho}_i$ used in Eqs. (\ref{omegai}) for the NG and pseudo-NG bosons, we
shall use $A_{\Omega}\tilde{\Omega}$ and $A_{\Omega}Re\tilde{\rho}$,
where $\tilde{\Omega}$ and $Re\tilde{\rho}$ are now column vectors.
With these definitions, one simply gets $\tilde{\Omega}^{\dagger}
Re\tilde{\rho}= \tilde{\Omega}^{\dagger}A_{\Omega}^{-1}
A_{\Omega} Re\tilde{\rho}$ . This simple assumption 
is used for two
purposes: 1) To reduce the number of arbitrary parameters; 2) To
see how far one can go with it before one needs to modify it.
With this assumption, the mass splitting among the pseudo-NG bosons
are given as in Eq. (\ref{Gmass2}), namely
\begin{equation}
M_{P3}^2 = M_P^2; M_{P1}^2 = M_P^2 (1 + x_1);  M_{P2}^2 = M_P^2 (1 + x_2),
\label{Pmass}
\end{equation}
with the same $x_i$ as for the gauge boson masses. Furthermore, the mixing angles
are the same as above. The contributions of
the ``rotated'' pseudo-NG bosons to the neutrino mass matrix elements
will therefore be accompanied by a factor 
$-\frac{\sin(2\beta)} {32\,\pi^2}$, just as in Eq. (\ref{replace}).

As mentioned above, in the full computation of the matrix elements, we
shall use, for convenience, $A_{ij}$ to denote the matrix elements of 
$A_{\Omega}$ instead of the representation of Eq. (\ref{Aomega}). 
One should then
recall that, because $A_{\Omega}$ is an orthogonal matrix, one has:
$\sum_{j} A_{ij}^2 =1$ and $\sum_{k} A_{ki} A_{kj} = 0$. The form
of the neutrino mass matrix elements will make use of these properties,
just as we have done above.

With the above remarks in mind, the full $4\times 4$ neutrino mass matrix is now
given by:
\begin{equation}
{\cal M}_{\nu}/m_N = \left(
\begin{array}{cccc}
m_{11}&m_{12}&m_{13}&0 \\
m_{12}&m_{22}&m_{23}&0 \\
m_{13}&m_{23}&m_{33}&0 \\
0&0&0& 1
\end{array}
\right)\
\label{numatrix}
\end{equation}
where $m_N$ is the mass of the 4th generation neutrino shown in
Eq. (\ref{m_N}). In Eq. (\ref{numatrix}), we have ignored the 
tiny one-loop contribution
to $m_{44} \equiv 1$, in particular when there is no mixing
between the 4th neutrino and the lighter three. As we shall see
later on, it can be ignored even if there is mixing, the
reason being the fact that $m_{ij}$, $i,j=1,2,3$, are so much smaller
than $m_{44} \equiv 1$. A change of $m_{44}$ to a value slightly less 
than or greater than one will not significantly affect the eigenvalues,
as we shall see in the numerical examples below.

With
\begin{equation}
\Delta I(G,P,x_i) \equiv I(M_G, x_i)- I(M_P, x_i),
\end{equation}
where $I(M_P, x_i)$ is given by Eq. (\ref{int2}) with the substitution
$M_G \rightarrow M_P$, one obtains for $m_{ij}$:
\begin{mathletters}
\label{mij}
\begin{equation}
m_{11}= \frac{\sin(2\beta)}{32\,\pi^2}\{\Delta I(G,P)-
A_{11}^2 x_1 \Delta I(G,P,x_1) - A_{12}^2 x_2 \Delta I(G,P,x_2)\} 
\end{equation}
\begin{equation}
m_{22}= \frac{\sin(2\beta)}{32\,\pi^2}\{\Delta I(G,P)-
A_{21}^2 x_1 \Delta I(G,P,x_1) - 
A_{22}^2 x_2 \Delta I(G,P,x_2)\} 
\end{equation}
\begin{equation}
m_{33}= \frac{\sin(2\beta)}{32\,\pi^2}\{\Delta I(G,P)-
A_{31}^2 x_1 \Delta I(G,P,x_1) - 
A_{32}^2 x_2 \Delta I(G,P,x_2)\} 
\end{equation}
\begin{equation}
m_{12}= -\frac{\sin(2\beta)}{32\,\pi^2}\{A_{11}
A_{21} x_1 \Delta I(G,P,x_1) + A_{12} A_{22} x_2 \Delta I(G,P,x_2)\} 
\end{equation}
\begin{equation}
m_{13}= -\frac{\sin(2\beta)}{32\,\pi^2}\{A_{11}
A_{31} x_1 \Delta I(G,P,x_1) + A_{11} A_{32} x_2 \Delta I(G,P,x_2)\} 
\end{equation}
\begin{equation}
m_{23}= -\frac{\sin(2\beta)}{32\,\pi^2}\{A_{21}
A_{31} x_1 \Delta I(G,P,x_1) + A_{22} A_{32} x_2 \Delta I(G,P,x_2)\} 
\end{equation}
\end{mathletters}
where $A_{ij}$ denote the matrix elements of $A_{\Omega}$, as mentioned
above, and where $\Delta I(G,P)$ was already defined in Eq. (\ref{DeltaGP}).

A few remarks are in order here. First,
one can see that, in the limit $x_i \rightarrow 0$, ${\cal M}_{\nu}$
reduces to a diagonal matrix with three equal diagonal elements:
$ \frac{\sin(2\beta)}{32\,\pi^2}\Delta I(G,P)$.
Secondly, apart from various mixing angles, the off-diagonal
elements depend on results of loop integrals, $\Delta I(G,P,x_i)$ 
which, in turns, depend on the same parameters
as the ones that enter the loop integrals of the diagonal elements
in the unbroken case, $\Delta I(G,P)$. The ratio
$R_{I} \equiv \Delta I(G,P,x_i)/\Delta I(G,P)$ is plotted in Figs. (6,7), for
two values of the parameter $x$, as a function of $M_2$
in the similar manner to Fig. 2-5. (The two values of $x$ were
chosen for the purpose of illustration and to coincide with the two
examples given below.)
It can be seen that
the ratio $R_{I}$ is at most of O($10^-2$), even for $x$ as
large as 0.5. Therefore, in our model,
a small mass splitting in the scalar and gauge sectors results in
a scenario with almost degenerate light neutrinos. The difference of 
the mass squared, $\Delta m^2$, depends, however, on the size of
the off-diagonal elements.
To see how it
actually works, a simple model of mixings will be presented below
along with some numerical examples.

We starts out with a very simplistic model of mixing and
try to see how far one can go. It is:
\begin{equation}
{\cal M}_{G,P}^2 = M_{G,P}^2\left(
\begin{array}{ccc}
1&b&0 \\
b&1&0 \\
0&0&1
\end{array}
\right)\
\label{GPmatrix}
\end{equation}
where $b$ is a small parameter less than unity. This simple model
has the merit of elucidating the points that we have made above. (An
extension of this model, showing similar results, will be discussed 
below.) The above mass mixing (\ref{GPmatrix}) could come, for example, 
from a term
in the Lagrangian of the form: $\lambda_{5}(
(\Omega^{\alpha} \rho^{\prime}_{\alpha})(\Omega^{\beta} 
\rho^{\prime\prime}_{\beta}) +
(\rho^{\alpha} \rho^{\prime}_{\alpha})(\rho^{\beta} 
\rho^{\prime\prime}_{\beta}))$. Assuming $<\rho^{\prime}> = 
(v^{\prime},0,0,0)$, $<\rho^{\prime\prime}> = (0,v^{\prime\prime},0,0)$, 
with $v^{\prime,\prime\prime} \ll V,V^{\prime}$, one can obtain the 
above mass mixing matrix.

It is easy to see that the eigenvalues of (\ref{GPmatrix}) are
\begin{equation}
M_{G1,P1}^2 =  M_{G,P}^2 (1+b),  M_{G2,P2}^2 =  M_{G,P}^2 (1-b),
M_{G3,P3}^2 =  M_{G,P}^2.
\end{equation}
$A_{\Omega}$ as discussed above is now given by
\begin{equation}
A_{\Omega} = \left(
\begin{array}{ccc}
\frac{1}{\sqrt{2}}&\frac{1}{\sqrt{2}}&0 \\
-\frac{1}{\sqrt{2}}&\frac{1}{\sqrt{2}}&0 \\
0&0&1
\end{array}
\right)\
\label{AOM}
\end{equation}
Now we can make the following identifications: $x_1 \equiv b$,
$x_2 \equiv -b$. The various angles are given in $A_{\Omega}$.
The matrix elements of the neutrino mass matrix are now 
fairly simple:
\begin{mathletters}
\label{mij2}
\begin{equation}
m_{11}= \frac{\sin(2\beta)}{32\,\pi^2}\{\Delta I(G,P)-
\frac{1}{2}b( \Delta I(G,P,b) - \Delta I(G,P,-b))\} 
\end{equation}
\begin{equation}
m_{22}= \frac{\sin(2\beta)}{32\,\pi^2}\{\Delta I(G,P)-
\frac{1}{2}b( \Delta I(G,P,b) - \Delta I(G,P,-b))\} 
\end{equation}
\begin{equation}
m_{33}= \frac{\sin(2\beta)}{32\,\pi^2}\{\Delta I(G,P)\}
\end{equation}
\begin{equation}
m_{12}= -\frac{\sin(2\beta)}{32\,\pi^2}\{
\frac{1}{2}b( \Delta I(G,P,b) + \Delta I(G,P,-b))\} 
\end{equation}
\begin{equation}
m_{13}= 0,
\end{equation}
\begin{equation}
m_{23}= 0 .
\end{equation}
\end{mathletters}

The above matrix elements are surprisingly easy to handle. When they are
substituted into Eq. (\ref{numatrix}), one obtains straightforwardly the following
mass eigenvalues:
\begin{mathletters}
\label{numass}
\begin{equation}
m_1 = m_N\frac{\sin(2\beta)}{32\,\pi^2}\{\Delta I(G,P)-
b\Delta I(G,P,b)\},
\end{equation}
\begin{equation}
m_2 = m_N \frac{\sin(2\beta)}{32\,\pi^2}\Delta I(G,P) ,
\end{equation}
\begin{equation}
m_3 = m_N \frac{\sin(2\beta)}{32\,\pi^2}\{\Delta I(G,P)+
b\Delta I(G,P,-b)\} ,
\end{equation}
\begin{equation}
m_4 = m_N
\end{equation}
\end{mathletters}
The matrix which diagonalizes the above neutrino mass matrix is simply
\begin{equation}
U_{\nu} = \left(
\begin{array}{cccc}
\frac{1}{\sqrt{2}}&\frac{1}{\sqrt{2}}&0&0 \\
0&0&1&0 \\
-\frac{1}{\sqrt{2}}&\frac{1}{\sqrt{2}}&0&0 \\
0&0&0&1
\end{array}
\right)\
\label{Unu}
\end{equation}

One obtains the
following mass splittings:
\begin{mathletters}
\label{masssplit}
\begin{equation}
m_3^2 - m_2^2 = ( m_N \frac{\sin(2\beta)}{32\,\pi^2})^2 
(2 b \Delta I(G,P) \Delta I(G,P,-b) +(b\Delta I(G,P,-b))^2)  ,
\end{equation}
\begin{equation}
m_2^2 - m_1^2 = ( m_N \frac{\sin(2\beta)}{32\,\pi^2})^2
( 2 b \Delta I(G,P) \Delta I(G,P,b) + (b\Delta I(G,P,b))^2)  .
\end{equation}
\end{mathletters}
In general, $\Delta I(G,P,x_i) \ll \Delta I(G,P)$,
and combined with the fact that $b < 1$, one has
$(b\Delta I(G,P,-b or b))^2 \ll 2 b \Delta I(G,P) \Delta I(G,P,-b or b)$.
One can then neglect the last terms in Eq.(\ref{masssplit}). Numerically, one has
$\Delta I(G,P,b) \approx \Delta I(G,P,-b)$. This implies that
$m_3^2 - m_2^2 \approx m_2^2 - m_1^2$, a quasi-degenerate mass splitting. This
holds for any value of $b$. Solar and atmospheric data suggest otherwise.
This necessitates the lifting of this quasi-degeneracy of the mass
splitting. To do this, we need to invoke some kind of mixing between
the 4th neutrino and the lighter three. In an indirect way, the
disparity between $\Delta m_{23}^2$ and $\Delta m_{21}^2$ indicates- in our
model- the influence of a 4th generation. Before discussing this issue which
will be presented in the next section, let us illustrate numerically a few
examples of the quasi-degenerate case.

First, a few useful points are in order here. Since $m_2 =
m_N \frac{\sin(2\beta)}{32\,\pi^2}\Delta I(G,P)$, one can rewrite the
above equations (\ref{masssplit}) as (neglecting the last terms on 
the right-hand side)
\begin{mathletters}
\label{masssplit2}
\begin{equation}
m_3^2 - m_2^2 = m_2 ( m_N 2 b \frac{\sin(2\beta)}{32\,\pi^2} 
\Delta I(G,P,-b))  ,
\end{equation}
\begin{equation}
m_2^2 - m_1^2 = m_2 ( m_N 2 b \frac{\sin(2\beta)}{32\,\pi^2}
 \Delta I(G,P,b))  .
\end{equation}
\end{mathletters}
For a fixed value of $m_2$, the size of the mass splitting, $\Delta m^2$,
depends on the size of the factor $m_N (2 b)\,(\sin(2\beta)/32\,\pi^2) 
\Delta I(G,P,-b\, or\, b)$. At first glance, it appears that one can obtain
$\Delta m^2$ to be as small as one wants with the appropriate choice of $b$.
Although it is true that it can be so, we will show that, $\Delta m^2$ 
can also be very small ($< 10^{-10} eV^2$), even when
$b \approx 1$. This depends on how large the masses of some of the
particles participating in the loop diagrams are. As a result,
by limiting $\Delta m^2 \geq 10^{-10} eV^2$, one puts a constraint on
those masses.

In Fig. 8, we present the ``median'' mass $m_2$ as a function of $M_2$
and $M_G$ for a given $M_P$ (as presented in Figs. (2-5)). The mass is
given in units of $(m_N/100 \,GeV)$. Similarly, we present in Figs.(9,10)
$m_3^2 - m_2^2$ and $m_2^2 - m_1^2$ as a function of the same masses,
but also for a given value of $b$. The results are expressed in units
of $(m_N/100\,GeV)^2$. For a more streamlined presentation of the
results, we shall limit ourselves to the case $m_2 \lesssim 1.67\,eV$,
coming from the suggestion that the sum of neutrino masses lies
between 4 and 5 $eV$ in order to form a component of HDM. Similarly,
we shall restrict $\Delta m^2 < 1\,eV^2$. In our model, for
a given value of $b$, $m_2$ and $\Delta m^2$ are correlated as one
can see from Fig. (8,9,10). 

Three major remarks are in order here.
1) One can see from Figs. (9, 10) the quasi-degeneracy of the mass
splitting in this particular scenario. (In the next section,
we shall see how one can lift that degeneracy.) 
2) One can
also see from Figs. (8, 9, 10) that, were the vacuum solution to the
solar neutrino problem favored, i.e. $\Delta m^2 \approx
10^{-10}\,eV^2$, the median value $m_2$ will
always be less than $0.1\,eV$. (The lifting of the mass
splitting degeneracy to satisfy the atmospheric neutrino data
will not change this conclusion.) This simply means that, at
least in this model, the solar vacuum solution is incompatible
with the light neutrinos being significant components of HDM.
3) Also from Figs. (8, 9, 10), it can be seen that the MSW solution,
$\Delta m^2 \approx 10^{-5}\,eV^2$, can correspond to values
of $m_2$ larger than 1 $eV$. (Again, the lifting of the mass
splitting degeneracy to satisfy the atmospheric neutrino data
will not change this conclusion.) So, in our scenario, the
MSW solution is compatible with the light neutrinos being 
significant components of HDM while the vacuum solution is not.
This is a very specific prediction of this model.

The above discussion leaves out the question of the size of
the mixing angles. As mentioned above, we  have already fixed the
neutrino mixing matrix $U_{\nu}$, as given by Eq. (\ref{Unu}). 
To complete the task, one has to model the charged lepton mixing matrix
$U_{l}$. This is something that we shall do in the last section.
We wish however to reemphasize the main result of this section: 
the values of $\Delta m^2$ are {\em independent} of $U_{l}$.
As one can see from Fig. (9, 10), $\Delta m^2$ depends only on the
various masses and on the parameter $b$, regardless of
$U_{l}$. As a consequence, the large angle or small angle
solutions as deduced from the data basically constrains,
in our scenario, the matrix $U_{l}$ ($U_{\nu}$ being already
fixed). 

To finish the discussion of this section, we wish to present
another form for the boson mass matrix, namely
\begin{equation}
{\cal M}_{G,P}^2 = M_{G,P}^2\left(
\begin{array}{ccc}
1&b&0 \\
b&1&b \\
0&b&1
\end{array}
\right)\
\end{equation}
The mass eigenvalues are
\begin{equation}
M_{G1,P1}^2 =  M_{G,P}^2 (1+\sqrt{2}b),  M_{G2,P2}^2 =  
M_{G,P}^2 (1-\sqrt{2}b),
M_{G3,P3}^2 =  M_{G,P}^2.
\end{equation}
$A_{\Omega}$ is now given by
\begin{equation}
A_{\Omega} = \left(
\begin{array}{ccc}
\frac{1}{2}&\frac{1}{\sqrt{2}}&\frac{1}{2} \\
\frac{1}{2}&-\frac{1}{\sqrt{2}}&\frac{1}{2} \\
-\frac{1}{\sqrt{2}}&0&\frac{1}{\sqrt{2}}
\end{array}
\right)\
\end{equation}
It is now straightforward to see that the neutrino mass matrix
elements are
\begin{equation}
m_{11}= m_{33} = \frac{\sin(2\beta)}{32\,\pi^2}\{\Delta I(G,P)-
\frac{b}{2\sqrt{2}}( \Delta I(G,P,b) - \Delta I(G,P,-b))\} 
\end{equation}
\begin{equation}
m_{22}= \frac{\sin(2\beta)}{32\,\pi^2}\{\Delta I(G,P)-
\frac{b}{\sqrt{2}}( \Delta I(G,P,b) - \Delta I(G,P,-b))\} 
\end{equation}
\begin{equation}
m_{12}= m_{23} = -\frac{\sin(2\beta)}{32\,\pi^2}\{
\frac{1}{2}b( \Delta I(G,P,b) + \Delta I(G,P,-b))\} 
\end{equation}
\begin{equation}
m_{13}= -\frac{\sin(2\beta)}{32\,\pi^2}\{
\frac{b}{2\sqrt{2}}( \Delta I(G,P,b) - \Delta I(G,P,-b))\} ,
\end{equation}
The eigenvalues are now simply given by
\begin{mathletters}
\begin{equation}
m_1 = m_N\frac{\sin(2\beta)}{32\,\pi^2}\{\Delta I(G,P)-
\sqrt{2}b\Delta I(G,P,b)\},
\end{equation}
\begin{equation}
m_2 = m_N \frac{\sin(2\beta)}{32\,\pi^2}\Delta I(G,P) ,
\end{equation}
\begin{equation}
m_3 = m_N \frac{\sin(2\beta)}{32\,\pi^2}\{\Delta I(G,P)+
\sqrt{2}b\Delta I(G,P,-b)\} ,
\end{equation}
\begin{equation}
m_4 = m_N
\end{equation}
\end{mathletters}
These masses have exactly the same form as those of Eq.(\ref{numass}), except
for the factor of $\sqrt{2}b$  instead of $b$. The matrix $U_{\nu}$
which diagonalizes the above matrix is exactly the same as in Eq. (\ref{Unu}).
Furthermore, $m_3^2 - m_2^2$ and $m_2^2 - m_1^2$ are of the same form as
Eqs. (\ref{masssplit}), with the following replacement in Eqs. (\ref{masssplit}): 
$b \rightarrow b^{\prime} = \sqrt{2} b$. The analysis which follows is 
exactly the same as the one presented above.

One can envision various scenarios for the boson mass matrices, but
it is certainly beyond the scope of this paper. To make things more
complicated than the simple assumption (\ref{GPmatrix}) does not 
appear to add much
to the discussion. Although it might be possible that a more involved
ansatz than (\ref{GPmatrix}) could lead to the lifting of the mass splitting
``quasi-degeneracy'', we have not succeeded in finding it. For this
reason, we now turn our attention to the more appealing scenario, 
at least within our model: the mixing between the 4th neutrino and
the rest.

\subsection{Neutrino mass matrix II: Mixing
between the 4th and the lighter three neutrinos }
\label{matrix2}

We have seen above that the simple ansatz for the boson mass matrices
(\ref{GPmatrix}) leads to a situation in which the mass splittings are
quasi-degenerate. This, of course, is in contradiction with
the data. In this model, in order to lift that quasi-degeneracy,
one needs a mixing between the 4th neutrino and at least one of the
lighter three. To get a feel for what might be needed, we shall first
present a few numerical examples. Based on these examples, we shall
attempt to give a theoretical basis for these numerical examples.

As an example, we shall choose a specific value for the parameter
$b$ and for the masses $M_2$, $M_G$, $M_P$ and $M_F$ which enter
the loop integrals for the neutrino masses. This will fix a definite
value for the matrix elements of the neutrino mass matrix. 
As we have already discussed earlier, the integrals depend 
only on the ratio of the above masses. We will present two examples
for the purpose of comparison. We shall see the reasons why we wish
to do so below.

1) First Example:

We shall set
(in units of $M_F$): $M_F=1$, $M_P=5$, $M_G = 10^6$, 
$M_2 = 2.5 \times 10^9$. For $b$, we shall choose: $b=0.035$.
(A smaller value of $b$ will give a smaller mass splitting.)
The reason for this choice (other choices are {\em equally valid}) is
the fact that it will give a typical mass of approximately
1.5 eV and a desired mass splitting. Putting these values
into the expressions for the integrals as given by Eq.(\ref{int2}), we
obtain the following neutrino mass matrix, where
$m_N$ is assumed to be 100 GeV for convenience:
\begin{equation}
{\cal M}_{\nu} = (100 GeV)\left(
\begin{array}{cccc}
-1.579332216 \times 10^{-11}&.8697647852 \times 10^{-17}&0&0 \\
.8697647852 \times 10^{-17}&-1.579332216 \times 10^{-11}&0&0 \\
0&0&-1.579332184 \times 10^{-11}&0 \\
0&0&0& 1
\end{array}
\right)\
\label{Mnu1}
\end{equation}
Notice that the above matrix has no mixing between the 4th
neutrino and the lighter three. The eigenvalues are just:
\begin{equation}
|m_1| = 1.579331346 eV; |m_2| =1.579332184 eV ; 
|m_3| =1.579333086 eV ; |m_4| = 100 GeV.
\end{equation}
As we have discussed in the previous section, this gives
a quasi-degenerate mass splitting,namely
\begin{equation}
\Delta m_{32}^2 = 1.601195367\times 10^{-6} eV^2,
\end{equation}
\begin{equation}
\Delta m_{21}^2= 1.535757079 \times 10^{-6} eV^2,
\end{equation}
where $\Delta m_{ji}^2 = m_{j}^2 - m_{i}^2$.

Let us now assume that the mixing with the 4th neutrino is
non-zero. We start out with the {\em simplest assumption}, namely
one in which only the 3rd neutrino mixes with the 4th one.
This means that $m_{34}$ and $m_{43}$ are both non-vanishing.
If we wish to have $m_3^2 - m_2^2 \approx 10^{-3} eV^2$ as
suggested by the atmospheric neutrino data, it turns out that
$m_{34}$ and $m_{43}$ cannot be too small nor too large,
being of order $10^{-7} m_N$. Notice that $m_{34}$ and $m_{43}$ do
not have to be equal.
We shall see how it might be
possible to obtain such a number. Let us first see how it works from a
numerical viewpoint.

To guide our understanding of how things work, let us notice that,
by adding $m_{34}$ and $m_{43}$ to ${\cal M}_{\nu}$ above, one
changes only one of the three light mass eigenvalues, leaving
the other two the same. Now the two unchanged eigenvalues
will be the ones that fix one of the two mass splittings,
$\Delta m^2$. For convenience, we shall choose the
$\Delta m^2$ corresponding to the unmodified mass eigenvalues as
the one which corresponds to the solar neutrino problem.
As we have learned from the above analysis
in Section (\ref{matrix1}), if one chooses the MSW solution, then one can find
masses which are large enough for HDM, while, if the vacuum solution
is chosen, the masses will be too small to form any significant
component of HDM. For the numerical example given here, we shall
choose the MSW solution as shown above. For $m_{34}$ and $m_{43}$,
we shall first choose a symmetric case (there is no particular reason
for this being so) as an example. We have
\begin{equation}
{\cal M}_{\nu} = (100 GeV)\left(
\begin{array}{cccc}
-1.579332216 \times 10^{-11}&.8697647852 \times 10^{-17}&0&0 \\
.8697647852 \times 10^{-17}&-1.579332216 \times 10^{-11}&0&0 \\
0&0&-1.579332184 \times 10^{-11}&.8 \times 10^{-7} \\
0&0&.8 \times 10^{-7}& 1
\end{array}
\right)\
\label{Mnu2}
\end{equation}
The eigenvalues are
\begin{equation}
|m_1| = 1.579331346 eV ; |m_2| = 1.579333086 eV ; |m_3| = 1.579972184 eV
; m_4 = 100 GeV.
\end{equation} 
We then get
\begin{equation}
\Delta m_{32}^2  = 2.02 \times 10^{-3} eV^2,
\label{m32}
\end{equation}
\begin{equation}
\Delta m_{21}^2  = 5.497 \times 10^{-6} eV^2.
\label{m21}
\end{equation}
The matrix which diagonalizes the mass matrix is
\begin{equation}
U_{\nu} = \left(
\begin{array}{cccc}
-\frac{1}{\sqrt{2}}&-\frac{1}{\sqrt{2}}&0&0 \\
\frac{1}{\sqrt{2}}&-\frac{1}{\sqrt{2}}&0&0 \\
0&0&-1&0.8 \times 10^{-7}\\
0&0&.8 \times 10^{-7}&1
\end{array}
\right)\
\label{Unu2}
\end{equation}

Two remarks are in order here. Firstly, from the values of the light
neutrino masses, one obtains $\sum_{i=1}^{3} |m_{i}| \approx 4.7 eV$,
which is in the range of mass for HDM. Secondly, Eq. (\ref{m32}) corresponds to
the best fit for the atmospheric neutrino data, while Eq. (\ref{m21})
corresponds to the best fit for the (small angle) MSW solution to the
solar neutrino data. One word of caution: this is not a prediction because
we chose the masses ($M_G$, $M_2$, etc...) in such a way as to
``reproduce'' the experimental results. It nevertheless shows a {\em dynamical}
basis for these numbers. Also, for nothing more than a numerical
example, the values of $m_{34,43}$ were chosen arbitrarily in order to
have the desired mass splitting. How to justify these values is the subject
to be discussed below.

The next numerical example deals with the case when
$m_{34}\neq m_{43}$. In doing the analysis, we notice that
it does not matter whether  $m_{34}$ is greater than $m_{43}$
or the other way around. One obtains the same result either
way. We shall require that $\Delta m_{32}^2(eV^2) =10^{-3}-10^{-2}$.
It turns out that $m_{34}$ and $m_{43}$ can range (in units of
$m_N$) only between approximately $0.4 \times 10^{-6}$ and
$0.8 \times 10^{-8}$. To be explicit, one has
\begin{equation}
{\cal M}_{\nu} = (100 GeV)\left(
\begin{array}{cccc}
-1.579332216 \times 10^{-11}&.8697647852 \times 10^{-17}&0&0 \\
.8697647852 \times 10^{-17}&-1.579332216 \times 10^{-11}&0&0 \\
0&0&-1.579332184 \times 10^{-11}&.4 \times 10^{-6} \\
0&0&.8 \times 10^{-7}& 1
\end{array}
\right)\
\end{equation}
gives $\Delta m_{32}^2(eV^2) \approx 10^{-2}$, while
\begin{equation}
{\cal M}_{\nu} = (100 GeV)\left(
\begin{array}{cccc}
-1.579332216 \times 10^{-11}&.8697647852 \times 10^{-17}&0&0 \\
.8697647852 \times 10^{-17}&-1.579332216 \times 10^{-11}&0&0 \\
0&0&-1.579332184 \times 10^{-11}&.4 \times 10^{-6} \\
0&0&.8 \times 10^{-8}& 1
\end{array}
\right)\
\end{equation}
gives $\Delta m_{32}^2(eV^2) \approx 10^{-3}$. Notice that
$\Delta m_{21}^2$ stays the same.
The above numerical results show that $m_{34}$ can differ
from $m_{43}$ by a large factor (50 in this case) while 
keeping $\Delta m_{32}$ within the desired range. 

2) Second Example:

In this example, we choose
(in units of $M_F$): $M_F=1$, $M_P=5$, $M_G = 10^4$, 
$M_2 = 1.2 \times 10^9$. For $b$, we shall choose: $b=0.000095$.
For simplicity, we shall assume, as we have already done above,
the following values for $m_{34,43}$, namely $m_{34} = m_{43} =
0.8 \times 10^{-7} (100 GeV)$. The mass matrix is now
\begin{equation}
{\cal M}_{\nu} = (100 GeV)\left(
\begin{array}{cccc}
1.382258467 \times 10^{-11}&.981382953 \times 10^{-17}&0&0 \\
.981382953 \times 10^{-17}&1.382258467 \times 10^{-11}&0&0 \\
0&0&1.382258467 \times 10^{-11}&.8 \times 10^{-7} \\
0&0&.8 \times 10^{-7}& 1
\end{array}
\right)\
\label{Mnu3}
\end{equation}
The eigenvalues are
\begin{equation}
m_1 = 1.382259448 eV; m_2 = 1.382257486 eV; m_3 = 1.381618467 eV;
m_4 = 100 GeV,
\end{equation}
with the corresponding diagonalization matrix given by
\begin{equation}
U_{\nu} = \left(
\begin{array}{cccc}
-\frac{1}{\sqrt{2}}&-\frac{1}{\sqrt{2}}&0&0 \\
\frac{1}{\sqrt{2}}&-\frac{1}{\sqrt{2}}&0&0 \\
0&0&-1&0.8 \times 10^{-7}\\
0&0&.8 \times 10^{-7}&1
\end{array}
\right)\
\label{Unu3}
\end{equation}
The mass splittings are
\begin{equation}
|\Delta m_{32}^2|  = 1.77 \times 10^{-3} eV^2,
\label{m32p}
\end{equation}
\begin{equation}
|\Delta m_{21}^2|  = 5.42 \times 10^{-6} eV^2.
\label{m21p}
\end{equation}

The above two examples are chosen sololy for illustration. Other
values of $\Delta m^2$ are possible with different choices
of various masses ($M_G$, $M_2$, etc..) and/or the parameter $b$.

Before turning to the discussion on the possible origins
of $m_{34,43}$, let us briefly discuss the ``tiny''
one-loop contribution to $m_{44}$, namely $\delta m_4$
as given by Eq. (\ref{delm4}). One might wonder how it would affect
the light mass eigenvalues.
It turns out however that, as long as 
$\delta m_4 \ll 1$ (which is the case in this paper), it
does not matter what value it takes. It is easy to see how.
A $2\times2$ matrix of the form $(a,c;c,b)$, where
$a,c \ll b$, has as eigenvalues: $b+c^2/b+(1/4)a^2/b)+O(c^4,a^4)$
and $a-c^2/b-(1/4)a^2/b)+O(c^4,a^4)$. One can see that, for the
smaller eigenvalue, a small change in $b$ affects very little its
value. As an example, we put 0.99 instead of 1 in Eq. (). We obtain
$\Delta m_{32}^2(eV^2) \approx 1.02 \times 10^{-3}$ instead
of $1.06 \times 10^{-3}$ (for 1). If we put 1.1 instead of 1,
we obtain $\Delta m_{32}^2(eV^2) \approx 0.923 \times 10^{-3}$.
Considering the kind of accuracy that one has at the present time,
this is completely irrelevant.

There are probably several scenarios for calculating
$m_{34,43}$. However, considering the fact that the present experimental 
status is not accurate enough for a detailed model, we will
present below a more or less ``generic'' scenario which will
show how one can obtain $m_{34,43}$ of the right order of
magnitude.

What might be the origin of $m_{34,43}$? It might be obvious up
until now that the vacuum expectation values of $\Omega$ and $\rho$
shown in Subsection (III.A) cannot generate such a mixing. One needs at least one
additional scalar with a non-vanishing vacuum expectation value
along the 3rd direction. Let that field be $\Omega^{\prime}$ and
let us assume that $<\Omega^{\prime}> = (0,0,\tilde{v},0)$. Let us also
assume that there are couplings of the type:
\begin{equation}
\lambda_{34}\Omega^{\alpha}\Omega^{\prime}_{\alpha}
\rho^{\beta} \rho_{\beta};
\lambda_{43}\Omega^{\alpha}\rho_{\alpha}
\Omega^{\prime,\beta}\rho_{\beta},
\end{equation}
where, for convenience, we have omitted the $SU(2)_{\nu R}$ index
in $\rho$. With the above couplings, one can construct diagrams
for $m_{34}$ and $m_{43}$ as shown in Fig. 11.

We shall denote the masses of $\tilde{H}_4$ and $\tilde{h}_4$ by
$M_{H_4}$ and $M_{h_4}$ respectively.
Let us define the following quantities:
\begin{mathletters}
\begin{equation}
\Delta M^2(G,\tilde{H}_4) = M_{G}^2 - M_{H_4}^2 ,
\end{equation}
\begin{equation}
\Delta M^2(G,\tilde{h}_4) = M_{G}^2 - M_{h_4}^2 ,
\end{equation}
\begin{equation}
\Delta M^2(P,\tilde{H}_4) = M_{P}^2 - M_{H_4}^2 ,
\end{equation}
\begin{equation}
\Delta M^2(P,\tilde{h}_4) = M_{P}^2 - M_{h_4}^2 .
\end{equation}
\end{mathletters}
From Fig. 8, we obtain:
\begin{mathletters}
\label{m3443}
\begin{eqnarray}
m_{34}/m_N &=& (\frac{\lambda_{34}}{16 \pi^2})(\frac{\tilde{v}M_F M_2}
{V})(c_{\beta}^2 s_{\alpha}^2 \frac{\Delta I(G,\tilde{H}_4)}
{\Delta M^2(G,\tilde{H}_4)} +c_{\beta}^2 c_{\alpha}^2 
\frac{\Delta I(G,\tilde{h}_4)}
{\Delta M^2(G,\tilde{h}_4)}+s_{\beta}^2 s_{\alpha}^2 
\frac{\Delta I(P,\tilde{H}_4)}
{\Delta M^2(P,\tilde{H}_4)} \nonumber \\
& &+s_{\beta}^2 c_{\alpha}^2 
\frac{\Delta I(G,\tilde{h}_4)}
{\Delta M^2(G,\tilde{h}_4)}),
\end{eqnarray}
\begin{eqnarray}
m_{43}/m_N& = &(\frac{\lambda_{34}}{16 \pi^2})(\frac{\tilde{v}M_F M_2}
{V})(s_{\beta}^2 c_{\alpha}^2 \frac{\Delta I(G,\tilde{H}_4)}
{\Delta M^2(G,\tilde{H}_4)} +s_{\beta}^2 s_{\alpha}^2 
\frac{\Delta I(G,\tilde{h}_4)}
{\Delta M^2(G,\tilde{h}_4)}+c_{\beta}^2 c_{\alpha}^2 
\frac{\Delta I(P,\tilde{H}_4)}
{\Delta M^2(P,\tilde{H}_4)} \nonumber \\
& &+c_{\beta}^2 s_{\alpha}^2 
\frac{\Delta I(G,\tilde{h}_4)}
{\Delta M^2(G,\tilde{h}_4)}),
\end{eqnarray}
\end{mathletters}
where $c$ and $s$ stand for $\cos$ and $\sin$, and $\Delta I(G, \tilde{H}_4)$
and the other similar quantities in Eq. (\ref{m3443}) are given 
by Eq. (\ref{DeltaGP}), with the
substitution of the appropriate masses taken into account.

As one can see from the above equations, the expressions appear rather
complicated at first look. However, one can make an estimate as to
which term in $m_{34}$ and $m_{43}$ is the most important. Each term
in Eqs. (\ref{m3443}) is of the form: $\lambda (\tilde{v}/V)(M_F/M_2)
(M_2^2/\Delta M^2)\Delta I (mixing\, angles)$, where $\lambda$
stands for $\lambda_{34,43}$. First, we have seen from the above
numerical analysis that, if we wish to have a mass of O(1-2 eV),
then $M_F/M_2 \approx 10^{-9}$. It is reasonable to assume that
$\lambda (\tilde{v}/V) \times (mixing\, angles) \leq 1$. If one of the terms
in Eq. (\ref{m3443}) were to be the dominant one and that $m_{34,43} \approx
10^{-7}$, then one should have $(M_2^2/\Delta M^2)\Delta I \gtrsim
10^2$. Let us first look at the $(G;\tilde{H},\tilde{h})$ contribution.
Assuming that $M_{H_4,h_4} < M_G$ so that $(M_2^2/\Delta M^2)
\approx M_2^{2}/M_G^2$, it turns out numerically that
$(M_2^{2}/M_G^2) \Delta I$ is always less than $\sim 10$.
For $M_{H_4,h_4} > M_G$, $\Delta I$ is larger in value than the
previous case, but then with $(M_2^2/\Delta M^2)
\approx M_2^{2}/M_{H_4,h_4}^2$, one will again have
$(M_2^{2}/M_{H_4,h_4}^2) \Delta I$ less than $10^2$. Taking
into account the actual calculation of $m_{34,43}$ which
includes mixing angles and various factors, the
$(G;\tilde{H},\tilde{h})$ would be too small to actually affect
the mass splittings. This leaves us with the contribution
coming from $(P;\tilde{H},\tilde{h})$. Here, as we have
done above, we will set $M_P = 5$ in units of $M_F$. There
are several possibilities that one can explore. We will
present here one of such possibilities. The main purpose
will be to show that, under reasonable assumptions, one can
obtain the desired order of magnitude for $m_{34,43}$. In
addition, one would like to see phenomenological implications
coming from such a scenario- something extra other than just
a mass matrix.

Let us assume that, by an appropriate choice of parameters
in the Higgs potential, one has $M_{H_4}$ to be of O($M_G$),
and that $M_{h_4} \ll M_2$. Furthermore, let us assume that
one also has $\beta \approx \alpha$. Although it is not 
really necessary, let us further assume that $\lambda_{34}
\sim \lambda_{43}$. Now numerically,
$(M_2^2/\Delta M^2)\Delta I < 10^2$ when one of the masses
in $\Delta M^2$ is much larger than the other one and not too
much different from $M_2$. This is just the case for $M_{H_4}
= O(M_G) \gg M_P$. Under these assumptions, we are left with
the $(P;h)$ contribution. In this case, one has
$m_{34} \approx m_{43}$. So we get
\begin{equation}
|m_{34}| \approx |m_{43}| \approx m_N \lambda_{34} \frac{\tilde{v}}{V}
\frac{M_F}{M_2}|\frac{M_2^2}{M_P^2-M_{h_4}^2}|\Delta M^2(G,\tilde{h}_4)
s_{\beta}^2 c_{\alpha}^2.
\label{m34}
\end{equation}
Typically, $\Delta M^2(G,\tilde{h}_4) = O(10^{-7}-10^{-11})$. In
most of our examples, $M_F/M_2 \sim 10^{-9}$. So one would expect
$(M_F/M_2)\Delta M^2(G,\tilde{h}_4) \sim 10^{-16}-10^{-20}$. If
we wish $m_{34} \approx m_{43} \approx m_N .8 \times 10^{-7}$, 
for example,
the other factors have to be sufficiently large. First, the
ratio $|\frac{M_2^2}{M_P^2-M_{h_4}^2}|$ can be rather large if
$M_{h_4}$ is small compared with $M_2$. Secondly, even if the
previous ratio can be large, it can still be offset by 
$s_{\beta}^2 c_{\alpha}^2$. Let us recall from Eq. (\ref{beta1}) that
$\tan \beta \approx g^2 (M_F/M_2)(M_2^2/M_G^2) \approx g^2
10^{-9} (M_2^2/M_G^2)$. Therefore the angle can be very small
if $M_G$ is too ``close'' in mass to $M_2$. A numerical 
investigation reveals that, if one wants to have a mass
of $O(1eV)$ and, at the same time, a large enough angle,
$M_G$ can be relatively ``low'' ($\sim 10^4$ in units of $M_F$).
(This would imply that the scale of family symmetry could be
a few thousands of TeV if $M_F$ is a few hundred GeV's.)
We now give a couple of numerical estimates. We shall take
the Second Example as a prototype. There one can calculate
the factor $s_{\beta}^2 c_{\alpha}^2$ to be $\approx 0.134$.
1) For $M_{h_4} = 100$ with all other masses being the same
as those of the Second Example, we obtain
\begin{equation}
m_{34} \approx m_{43} \approx  m_N \lambda_{34} \frac{\tilde{v}}{V}
\times 4.7 \times 10^{-7}.
\end{equation}
If we wish $m_{34} \approx m_{43} \approx m_N .8 \times 10^{-7}$,
then $\lambda_{34} \frac{\tilde{v}}{V} \approx 0.17$. So one
could either have $\lambda_{34} \approx .2$ and $\tilde{v} \approx V$,
or some other combination.
2)For $M_{h_4} = 10$, we have 
\begin{equation}
m_{34} \approx m_{43} \approx  m_N \lambda_{34} \frac{\tilde{v}}{V}
\times 1.4 \times 10^{-5},
\end{equation}
which would imply $\lambda_{34} \frac{\tilde{v}}{V} \approx 0.006$-
a reasonable constraint. 

It turns out that the cases with $M_{h_4} \geq 1000$ (in units of
$M_F$) do not work
because then the mass ratios are not large enough to compensate
for the smallness of the integrals. It is interesting that one
can have scenarios where $\tilde{h}_4$ is light enough (i.e.
not too much heavier than $F$)- a feature which could have
interesting phenomenological implications.

\subsection{Oscillation Angles}

To discuss the neutrino oscillation angles, one needs to give the
leptonic ``CKM'' matrix, namely $V_L = U_{l}^{\dagger}U_{\nu}$.
It is beyond the scope of this paper to discuss the charged lepton
sector, and hence $U_{l}$. This will be the subject of the following
publication. However, we can give an example of $U_{l}$
by adopting, at least for this paper, a simple model of charged
lepton masses of Ref. (\cite{yanagida}), which is a phenomenological model
based on a generalization to the lepton sector of the ``democratic
mass'' ansatz of the quark sector. The reason why we use, as
an example, Ref. (\cite{yanagida}) is because the matrix which diagonalizes the
neutrino mass matrix, $U_{\nu}$, is  {\em identical} to 
the $3 \times 3$ submatrix of our Eq. (\ref{Unu}) (apart from a difference 
in in the overall sign), namely
\begin{equation}
U_{\nu}^{(3)} = \left(
\begin{array}{ccc}
-\frac{1}{\sqrt{2}}&-\frac{1}{\sqrt{2}}&0 \\
\frac{1}{\sqrt{2}}&-\frac{1}{\sqrt{2}}&0 \\
0&0&-1\\
\end{array}
\right)\
\label{Unu4}
\end{equation}
Although Ref. (\cite{yanagida}) discussed an ansatz for three generations, we
will use it here because the mixing with the 4th generation is
not relevant for the oscillation angles we are interested in. (It
was relevant for the mass splitting.) So, basically, we will be 
using {\em only} the phenomenological ansatz for the {\em charged
lepton} mass matrix of Ref. (\cite{yanagida}). In fact, we will {\em only} use the
matrix which diagonalizes that mass matrix.

The $3\times 3$ leptonic ``CKM'' matrix written down by Ref. (\cite{yanagida}) is
\begin{equation}
V_l = (AB_l)^{\dag}U_{\nu} \approx \left(
\begin{array}{ccc}
1&-(1/\sqrt{3})\sqrt{m_{e}/m_{\mu}}&(2/\sqrt{6})\sqrt{m_e/m_{\mu}} \\
\sqrt{m_e/m_{\mu}}&1/\sqrt{3}&-2/\sqrt{6} \\
0&2/\sqrt{6}&1/\sqrt{3}\\
\end{array}
\right)\
\label{Vl}
\end{equation}
where $AB_l$ is the matrix which diagonalizes the charged lepton mass 
matrix, $U_{\nu}$ is given above, and $m_e$ and $m_{\mu}$ are the 
electron and muon masses respectively. Now, the probability for
$\nu_{e} \rightarrow \nu_{\mu}$ is
\begin{equation}
P(\nu_{e} \rightarrow \nu_{\mu}) \approx 2(V_{11}^2V_{21}^2
+V_{12}^2V_{22}^2-V_{13}^2V_{23}^2)\sin^2(1.27\Delta m_{12}^2 L/E),
\label{nutonumu}
\end{equation}
where the usual notation $\sin^2(2\theta_{12})$ is simply the
coefficient of $\sin^2(1.27\Delta m_{12}^2 L/E)$. Similarly
\begin{equation}
P(\nu_{\mu} \rightarrow \nu_{\tau}) \approx 4V_{23}^2V_{33}^2
\sin^2(1.27\Delta m_{23}^2 L/E),
\label{numutonutau}
\end{equation}
with $\sin^2(2\theta_{23})$ being the coefficient of
$\sin^2(1.27\Delta m_{23}^2 L/E)$. Putting in the values
of $m_e$ and $m_{\mu}$ to evaluate the matrix elements of
$V_l$, one readily obtains
\begin{equation}
\sin^2(2\theta_{12}) \approx 6.5 \times 10^{-3};
\sin^2(2\theta_{23}) \approx 0.89.
\end{equation}
These results correspond to the {\em small angle} MSW
solution, and to the {\em large angle} atmospheric solution
respectively. This is consistent with the best fit for
the two neutrino oscillation problems.

The above results should be viewed with caution. The {\em small angle}
MSW solution given above, as well as the {\em large angle} solution
for the atmospheric oscillation, depends on the charged lepton sector
- the neutrino sector diagonalizatin matrix being already fixed by
Eq. (\ref{Unu}). One can easily imagine how these angles can drastically
change if the charged lepton mass matrix has a different texture.
This will be the subject of a subsequent paper where we will
examine the charged lepton mass matrix in the context of the present
model- the basic interaction Lagrangian being already given by Eq. (\ref{lag1}).

\section{Epilogue}

The above discussions focused entirely on the atmospheric and solar
neutrino data. We have left out the LSND \cite{LSND} result for
two reasons. Firstly, it is because it might be prudent to wait
for future experiments, either to confirm or to refute these results.
Secondly, it is because it is extremely hard to incorporate {\em all 
three} experiments simultaneously in a ``natural'' model. In general,
one needs to invoke some kind of sterile neutrino that mixes with
the lightest neutrino to explain the solar data. If this sterile
neutrino were to arise from some kind of model, it is rather 
hard to invent, in a ``natural'' way, a scenario to 
explain why this sterile neutrino is
so light and close in mass to one of the three active light neutrinos.

Let us suppose that the LSND result are verified by future experiments.
What does the model presented in this paper have to say about a
sterile neutrino? Let us remember that $\eta_R =(\nu_R^{\alpha}, 
\tilde{\nu}_R^{\alpha})$ is an electroweak {\em singlet}. Furthermore
we have seen that it is $\nu_R^{\alpha}$ which mixes with $l^{\alpha}_L$
to give masses to the neutrinos. Its $SU(2)_{\nu_R}$ partner,
$\tilde{\nu}_R^{\alpha})$, remains massless, at least within the
framework of the preceding sections. Could these be the so-called
sterile neutrinos? If so, how would they get a mass? How would they
mix with the light neutrinos? These are the questions which are
under investigation.

We have concentrated in this manuscript on the even option.
One might wonder about the odd option and its implication on
neutrino masses. It is beyond the scope of this paper to
investigate this issue, however
a preliminary investigation of the odd option, with three families
and one family singlet $\eta^{\prime}$, appears to indicate that
the preferred solution for the neutrino masses is that in which there
is a hierarchy $m_1 \ll m_2 \ll m_3$.

There are numerous phenomenological consequences to be worked out in
subsequent publications. One can, however, make one rather solid
prediction: neutrinos, being of the Dirac nature, will not give
rise to the phenomenon of neutrinoless double beta decay.
Another interesting consequence is the possible existence of ``light''
(i.e. 200 GeV or so) vector-like fermions: $F$, as well as TeV-scale
pseudo NG bosons which carry family and $SU(2)_{\nu_R}$ quatum
numbers. This will be dealt with in a separate paper.

Several other phenomenological issues remain to be investigated. 
For instance, what are the consequences
of a broken $SU(2)_{\nu R}$ and what might the cosmological implications of
$\tilde{\nu}_R$'s and $\eta^{\prime}_R$ be? When
$SU(2)_{\nu R}$ is broken by $\rho^{\alpha}_{i}$, the gauge bosons
are expected to acquire a mass of O($V^{\prime}$) and can be quite heavy. 
Since only {\em right-handed} neutral leptons participate in $SU(2)_{\nu R}$
interactions, a place where the effects of those gauge bosons might show up
is in the decays of neutrinos. Without going into detail, it is easy to see
that the decay of the light (near-degenerate) neutrinos into each other is
completely negligible for lack of phase space and for the fact that neutrino
masses are {\em tiny} compared with $V^{\prime}$ (even if the latter is in the TeV region).
This leaves us with the decay of the (heavy) fourth-generation neutral
lepton $N$ for which we have $N \rightarrow \tilde{N} + \nu_i + \tilde{\nu}_i$ (1)
via the exchange of $SU(2)_{\nu R}$ gauge bosons, and
$N \rightarrow l^{-}_i + l^{+}_j + \nu_j$ (2) if $m_N < m_W$ or
$N \rightarrow l^{-}_i + W$ (3) if  $m_N > m_W$. In addition, one could have
$N \rightarrow E + l^{+}_j + \nu_j$ when $m_N >m_E$, via the exchange
of $W$. Whether or not $m_N$ is larger or smaller than $m_E$, the
relevant decays to compare with each other are (1) and (3). To make an estimate,
let us assume the the family gauge coupling is about the same size as the electroweak
coupling ($g \sim 0.7$). The ratio of the decay widths for (1) and (3) is
approximately $\Gamma (1) /\Gamma (3) \sim 7.5 \times 10^{-4} 
(M_W/M_{\tilde{G}})^2 (m_N/M_{\tilde{G}})^2 (1- (M_W/m_N)^4)^{-2} x^{-2}$, where 
$M_{\tilde{G}}$ represent the mass of the $SU(2)_{\nu_R}$ gauge bosons
and $x$ represents the mixing cofficient between the 4th neutrino and a light charged
lepton. Now let us remember that the computation of the neutrino masses does
not involve $M_{\tilde{G}}$ and as a result there appears to be no constraint
there. However, $M_{\tilde{G}} \sim g V^{\prime}$ and $M_G \sim g V$, and
as a result $\tan \beta \equiv V^{\prime}/V \sim M_{\tilde{G}}/M_G
\sim g^2 10^{-9} (M_2^2/M_G^2)$. In the second example discussed in the previous
example, $V^{\prime} \sim V$ (with $M_G = 10^{4} M_F$) which implies
$M_{\tilde{G}} \sim M_G$. Now $\Gamma (1) /\Gamma (3)$ can also be appreciable if
$m_N$ is close to $m_W$. For example, if $M_F \sim 200 GeV$ and $m_N \sim 82 GeV$,
$\Gamma (1) /\Gamma (3) \sim 1$ provided $x \sim 10^{-9}$. If this were the case,
the signal would be quite interesting: a long-lived massive neutral lepton
whose electroweak decay width is not what it should be. It is certainly beyond
the scope of this paper to explore numerous phenomenological 
consequences which might arise from our scenario.

As for the cosmological
consequences of $\tilde{\nu}_R$'s and $\eta^{\prime}_R$, if they are massless,
one should recall our earlier discussion: These particles {\em only}
have family and $SU(2)_{\nu R}$ gauge interactions (both for $\tilde{\nu_R}$'s
and the latter only for $\eta^{\prime}_R$). Therefore, they {\em cannot} influence
big-bang nucleosynthesis. One can estimate their decoupling temperatures
by comparing the interaction rate $\Gamma_{int} \sim G^2 T^5$, where
$G^2 \sim 1/ (64 V^{(\prime) 4})$, with the Hubble rate $H \sim T^2/ m_{pl}$.
Decoupling occurs when $\Gamma_{int} < H$ which gives a temperature
of O($10^6$) GeV if $V^{(\prime)} \sim 10^9$ GeV for example. 
After this, their temperature would scale like $T \sim 1/R$. It
is not clear what else they can do except to exist as almost non-
interacting relativistic relics with an energy density negligible
compared with normal matter. At this stage, it is also not clear if
they really do need to have a mass. The cosmology of these objects
is probably worth exploring further.

Another interesting cosmological subject to explore is the ``heaviest''
particle in our scenario: The vector-like neutral fermion ${\cal M}_2$ which 
is singlet
under all the listed gauge groups in Eq. (\ref{gaugegroup}). $\cal{M}$
couples to other fermions via  \ref{lag1}. The decay modes obtained
from (\ref{lag1}) are: ${\cal M}_{2R} \rightarrow \phi^{\pm} F_{L}^{\mp}$ (1)
and ${\cal M}_{2L} \rightarrow \rho_{\alpha} \eta^{\alpha}$ (2). Notice that,
in the examples given above for the calculations of the neutrino masses,
the mass of this fermion is typically $M_2 \sim 10^{9} M_F$. So, if
$M_F \sim 200 \,GeV$ (or a few hundred GeV), one would then
expect the mass of ${\cal M}_{2}$ to be around  a few times 
$10^{11}$ GeV. If $M_F \sim 1 \,TeV$, ${\cal M}_{2}$ would have a mass
around $10^{12}$ GeV. 
The questions
that we would like to investigate are: (1) How many $\cal{M}_{2}$ are left
in the present universe?;(2) Could the decay of the relic $\cal{M}_2$'s
manifest itself as ultra high energy cosmic rays (UHECR)(with energy exceeding
$10^{20}\,eV = 10^{11}\,GeV$) whose origins are still unknown? It does
appear that the mass of ${\cal M}_{2}$ is in the right energy ballpark.
This would be the case of a non-accelerated source of UHECR and is part
of the ``top-down'' approach to UHECR \cite{UHECR}.
For example, ${\cal M}_{2R}$ would decay into the longitudinal component
of $W$ ($\phi^{\pm}$) and $ F_{L}^{\mp}$. $\phi^{\pm}$ would in turn decay
into extremely high-energy quarks and leptons. The quarks will hadronize
into hadrons such as pions which will eventually convert into photons,
neutrinos, etc..

Last but not least, in the subsequent series of papers, we shall deal
with the charged lepton sector and with the quark sector. In particular,
we shall see how the generalization of Eq. (\ref{lag1}) to the quark sector
might yield interesting results.



I would like to thank Vernon Barger, Paul Fishbane and Paul Frampton
for reading the manuscript and for useful comments.
This work is supported in parts by the US Department
of Energy under grant No. DE-A505-89ER40518.

\appendix
\section*{Higgs Potential}

In this Appendix, we shall discuss a simple form of the Higgs potential
for the group $SO(4) \otimes SU(2)_{\nu_R}$. For simplicity, we shall
assume that there is no cross coupling between ($\Omega$, $\rho$) and the
SM Higgs field $\phi$. (One might wonder about the fact that even if
the cross coupling were vanishing, it might still be induced through
radiative corrections. This, however, would be very small in our model.)

The potential containing $\Omega$ and $\rho$ reads
\begin{eqnarray}
V(\Omega, \rho)& =& \lambda_{1} (\Omega^{\alpha} \Omega_{\alpha} - V^2)^2
+ \lambda_{2} (\rho^{\dagger\, \alpha} \rho_{\alpha} - V^{\prime\, 2})^2
+\lambda_{3} [(\Omega^{\alpha} \Omega_{\alpha} - V^2) -
(\rho^{\dagger\, \alpha} \rho_{\alpha} - V^{\prime\, 2})]^2 \nonumber \\
& &+\lambda_{4} [(\Omega^{\alpha} \Omega_{\alpha})
(\rho^{\dagger\, \beta} \rho_{\beta})
-(\Omega^{\alpha} \rho^{\dagger}_{\alpha})(\Omega^{\beta} \rho_{\beta})
\label{potential}
\end{eqnarray}
where $<\Omega> = (0,0,0,V)$ and $<\rho> = (0,0,0,V^{\prime} \otimes s_1)$,
with $s_1 = \left( \begin{array}{c} 1 \\ 0 \end{array} \right)$.
Here, we will assume that $\Omega$ is real and $\rho$ is complex.
We will be particularly interested in the mass eigenstates resulting
from Eq. (\ref{potential}).

With $\Omega_4 = H_4 +V$ and $\rho_4 = \left( \begin{array}{c}
h_4 + V^{\prime} + i \phi_4 \\ \rho^{\prime}_4 \end{array} \right)$,
Eq. (\ref{potential}) gives rise to the following mass matrix
for $H_4$ and $h_4$:
\begin{equation}
8 \left( \begin{array}{cc}
(\lambda_1 + \lambda_3)V^2& -\lambda_3 V V^{\prime} \nonumber \\
 -\lambda_3 V V^{\prime}&(\lambda_2 + \lambda_3)V^{\prime 2}
\end{array}
\right)
\label{H4h4}
\end{equation} 
The eigenvectors are
\begin{mathletters}
\label{Hheigenvector}
\begin{equation}
\tilde{H}_4 = \cos \alpha\, H_4 + \sin \alpha\, h_4
\end{equation}
\begin{equation}
\tilde{h}_4 = -\sin \alpha\, H_4 + \cos \alpha\, h_4
\end{equation}
\end{mathletters}
The associated eigenvalues are
\begin{mathletters}
\label{physhiggs}
\begin{equation}
m^{2}_{H_4} = 4(\lambda_2 + \lambda_3) V^{\prime 2}m_{1}^2,
\end{equation}
\begin{equation}
m^{2}_{h_4} = 4(\lambda_2 + \lambda_3) V^{\prime 2}m_{2}^2 ,
\end{equation}
\end{mathletters}
where
\begin{mathletters}
\begin{equation}
m^{2}_{1,2} = \frac{1+a\pm \sqrt{(1-a)^2+4b^2}}{2},
\end{equation}
\begin{equation}
a = (\frac{\lambda_1 + \lambda_3}{\lambda_2 + \lambda_3}) \tan^2 \beta,
\end{equation}
\begin{equation}
b = (\frac{\lambda_3}{\lambda_2 + \lambda_3})\tan \beta,
\end{equation}
\begin{equation}
\tan \beta = \frac{V^{\prime}}{V},
\end{equation}
\begin{equation}
\cos \alpha = \frac{1}{\sqrt{1+[(1-m^{2}_{1})/b]^2}}.
\end{equation}
\end{mathletters}

The mass matrix for $\Omega_i$ and $Re\rho_i$ with $i=1,2,3$, is
\begin{equation}
2 \lambda_{4} \left(\begin{array}{cc}
V^{\prime 2}& -V V^{\prime} \nonumber \\
-V V^{\prime}& V^2
\end{array}
\right)
\label{OmegaRerho}
\end{equation}
The eigenvectors are
\begin{mathletters}
\label{OmRhoeigenvector}
\begin{equation}
\tilde{\Omega}_{i} = \cos \beta\, \Omega_{i} + \sin \beta\, Re\rho_{i},
\end{equation}
\begin{equation}
Re \tilde{\rho}_{i} = -\sin \beta\, \Omega_{i} + \cos \beta\, Re\rho_{i},
\end{equation}
\end{mathletters}
The associated eigenvalues are
\begin{mathletters}
\label{Omegarhomass}
\begin{equation}
m_{\tilde{\Omega}} = 0, 
\end{equation}
\begin{equation}
m_{Re\tilde{\rho}} = 2 \lambda_{4}
(V^{2} + V^{\prime 2}).
\end{equation}
\end{mathletters}
Notice that $\tilde{\Omega}_i$ are NG Goldstone bosons which are absorbed
by some of the $SO(4)$ gauge bosons. 

Since it is not of immediate relevance to the paper, we will simply
quote the masses of the other scalars obtained from (\ref{potential}).
Scalars (Pseudo-NG bosons) which have a mass $2\,\lambda_{4} V^2$: 
$Im\rho_i$, $Re\rho^{\prime}_{i}$,
$Im\rho^{\prime}_{i}$. Goldstone bosons which are absorbed by some of the
$SO(4) \otimes SU(2)_{\nu_R}$ gauge bosons: $Re\rho^{\prime}_{4}$,
$Im\rho^{\prime}_{4}$. Notice that the pseudo-NG boson masses are
all proportional to $\lambda_{4}$. As a result, their masses tend to
zero as $\lambda_{4} \rightarrow 0$.

\begin{figure}
\caption{Feynman graph showing the computation of $\tilde{G}_{\nu}$,
where $m_{\nu} = \tilde{G}_{\nu} \frac{v}{\sqrt{2}}$}
\end{figure}

\begin{figure}
\caption{The ratio $m_{\nu}/m_N$ (Eq. (23)) as a function of $M_2$
(in units of $M_F$, and hence the notation $M_F =1$), for $M_P =5$
and for various values of $M_G$. For visibility purpose, a few
curves have been inflated by factors $\times 10^{2,3,5,6}$.}
\end{figure}

\begin{figure}
\caption{The ratio $m_{\nu}/m_N$ (Eq. (23)) as a function of $M_2$
(in units of $M_F$, and hence the notation $M_F =1$), for $M_P =50$
and for various values of $M_G$. For visibility purpose, a few
curves have been inflated by factors $\times 10^{2,3,5,6}$.}
\end{figure}

\begin{figure}
\caption{The ratio $m_{\nu}/m_N$ (Eq. (23)) as a function of $M_2$
(in units of $M_F$, and hence the notation $M_F =1$), for $M_P =500$
and for various values of $M_G$. For visibility purpose, a few
curves have been inflated by factors $\times 10^{2,3,5,6}$.}
\end{figure}

\begin{figure}
\caption{The ratio $m_{\nu}/m_N$ (Eq. (23)) as a function of $M_2$
(in units of $M_F$, and hence the notation $M_F =1$), for $M_P =5000$
and for various values of $M_G$. For visibility purpose, a few
curves have been inflated by factors $\times 10^{2,3,5,6}$.}
\end{figure}

\begin{figure}
\caption{The ratio $R_{I} \equiv \Delta I(G,P,x_i)/\Delta I(G,P)$ for
$b=0.035$}
\end{figure}

\begin{figure}
\caption{The ratio $R_{I} \equiv \Delta I(G,P,x_i)/\Delta I(G,P)$ for
$b=0.000095$}
\end{figure}

\begin{figure}
\caption{The median mass $m_2$ as defined by Eq. (49b). Notice the 
correlation with $m_{2}^2 - m_{1}^2$ and $m_3^2 - m_2^2$ shown
in the next two figures }
\end{figure}

\begin{figure}
\caption{$m_{3}^2 - m_{2}^2$ as defined by Eq. (52a) for $b=0.000095$}
\end{figure}

\begin{figure}
\caption{$m_{2}^2 - m_{1}^2$ as defined by Eq. (52b) for $b=0.000095$}
\end{figure}

\begin{figure}
\caption{Diagram for $m_{34,43}$}
\end{figure}

\begin{table}
\caption{Particle content and quantum numbers of
$SU(3)_c \otimes SU(2)_L \otimes U(1)_Y \otimes SO(N_f) \otimes SU(2)_{\nu_R}$}
\begin{tabular}{l|r}
Standard Fermions      & $q_L = (3, 2, 1/6, N_f, 1)$\\
                       & $l_L = (1, 2, -1/2, N_f, 1)$\\
                       & $u_R = (3, 1, 2/3, N_f, 1)$\\
                       & $d_R = (3, 1, -1/3, N_f, 1)$\\
                       & $e_R = (1, 1, -1, N_f, 1)$\\ \hline
Right-handed $\nu$'s   & Option 1: $\eta_R = (1, 1, 0, N_f, 2)$ \\
                       & Option 2: $\eta_R = (1, 1, 0, N_f, 2)$;  \\ 
                       & $\eta_R^{\prime} = (1, 1, 0, 1, 2)$ \\ \hline
Vector-like Fermions   & $F_{L,R} = (1, 2, -1/2, 1, 1)$\\ 
                       & ${\cal M}_{1 L,R} = (1, 1, -1, 1, 1)$\\
                       & ${\cal M}_{2 L,R} = (1, 1, 0, 1, 1)$\\ \hline
Scalars                & $\Omega^{\alpha} = (1, 1, 0, N_f, 1)$\\
                       & $\rho_{i}^{\alpha} = (1, 1, 0, N_f, 2)$ \\
                       & $\phi = (1, 2, 1/2, 1, 1)$ 
\end{tabular}
\end{table}


\begin{references}
\bibitem[*]{email} e-mail: pqh@virginia.edu .
\bibitem{superk} Super-Kamiokande collaboration, Phys. Rev. Lett. {\bf 81},
1562 (1998).
\bibitem{kayser} See Peter Fisher, Boris Kayser, and Kevin S. McFarland, hep-
ph/9906244, for a comprehensive review and a list of references.
\bibitem{seesaw} M. Gell-Mann, P. Ramond, and R. Slansky in Supergravity,
edited by D. Freedman et al. (1979).
\bibitem{mohapatra1} J. C. Pati and A. Salam, Phys. Rev. D {\bf 10}, 275
(1974); R. N. Mohapatra and J. C. Pati, {\em ibid.} {\bf 11}, 2558 (1975);
R. N. Mohapatra and G. Sejanov\'{i}c, Phys. Rev. Lett. {\bf 44}, 912 (1980);
Phys. Rev. D {\bf 23}, 165 (1981), and references therein.
\bibitem{hung} P. Q. Hung,  Phys. Rev. {\bf D59}, 113008 (1999).
\bibitem{caldwell} See David O. Caldwell, hep-ph/9910349, for a review
of the subject.
\bibitem{witten} E. Witten, Phys. Lett. {\bf B117}, 324 (1982).
\bibitem{silva-marcos} J. I. Silva-Marcos, Phys. Rev. {\bf D59}, 091301
(1999).
\bibitem{erler} See Jens Erler, hep-ph/9903449, for a recent review and
a list of references therein.
\bibitem{hung1} P. Q. Hung, Phys. Rev. Lett. {\bf 80}, 3000 (1998).
\bibitem{hung2} P. H. Frampton and P. Q. Hung, Phys. Rev. D {\bf 58}, 057704 (1998).
\bibitem{physrep} P. H. Frampton, P. Q. Hung, and M. Sher, hep-ph/9903387,
Physics Reports in press.
\bibitem{2beta} Heidelberg-Moscow Collaboration, Phys. Rev. Lett. {\bf 83}, 41 (1999).
\bibitem{LEP} B. Adeva {\em et al} (L3); D. Decamp {\em et al} (ALEPH);
M. Z. Akraway {\em et al} (OPAL); P. Aarinio {\em et al} (DELPHI),
Phys. Lett. {\bf B231}, 509 (L3), 519 (ALEPH), 530 (OPAL), 539 (DELPHI) (1989).
\bibitem{yanagida} M. Fukugita, M. Tanimoto, T. Yanagida, Phys. Rev. {\bf
D57}, 4429 (1998).
\bibitem{LSND} LSND collaboration, Phys. Rev. Lett. {\bf 81}, 1774 (1998).
\bibitem{UHECR} For a comprehensive review, see 
P. Bhattacharjee and G. Sigl, astro-ph/9811011, Physics
Reports in press.
\end{references}
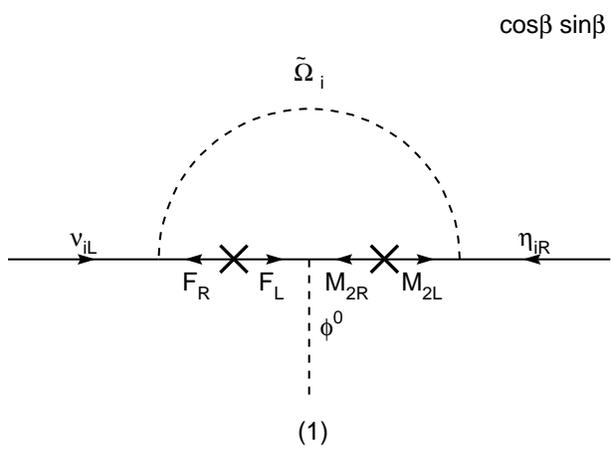
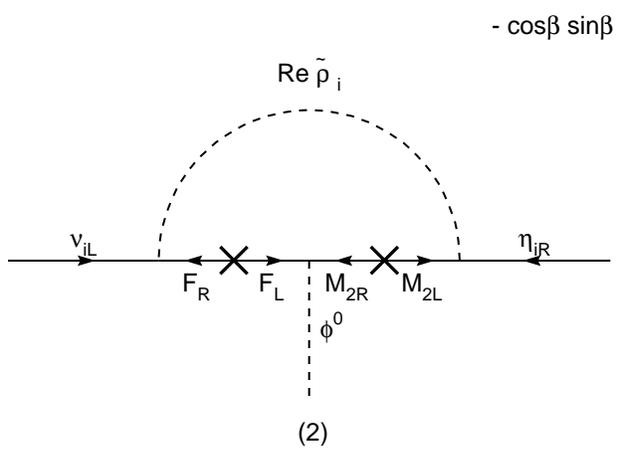
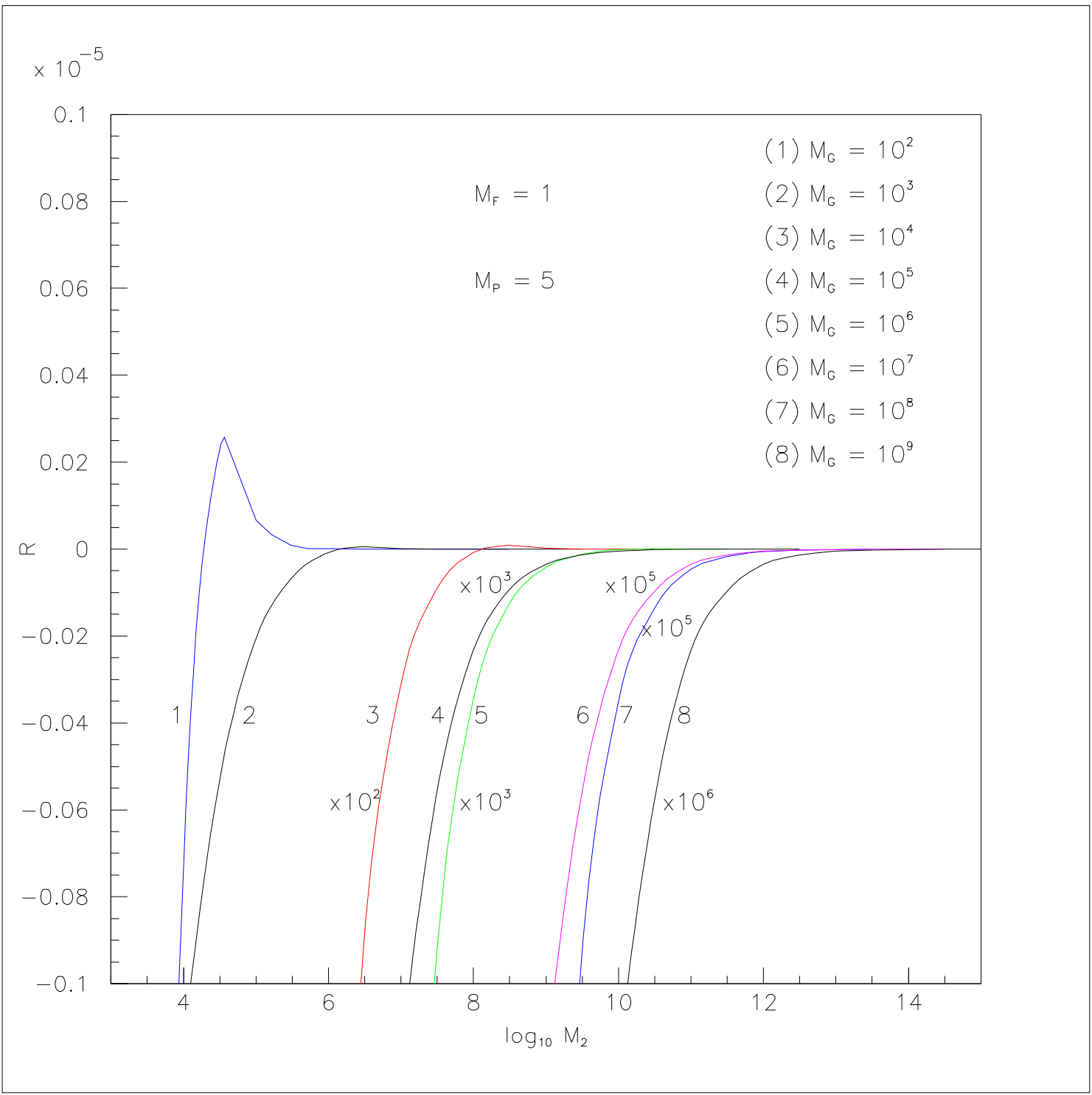
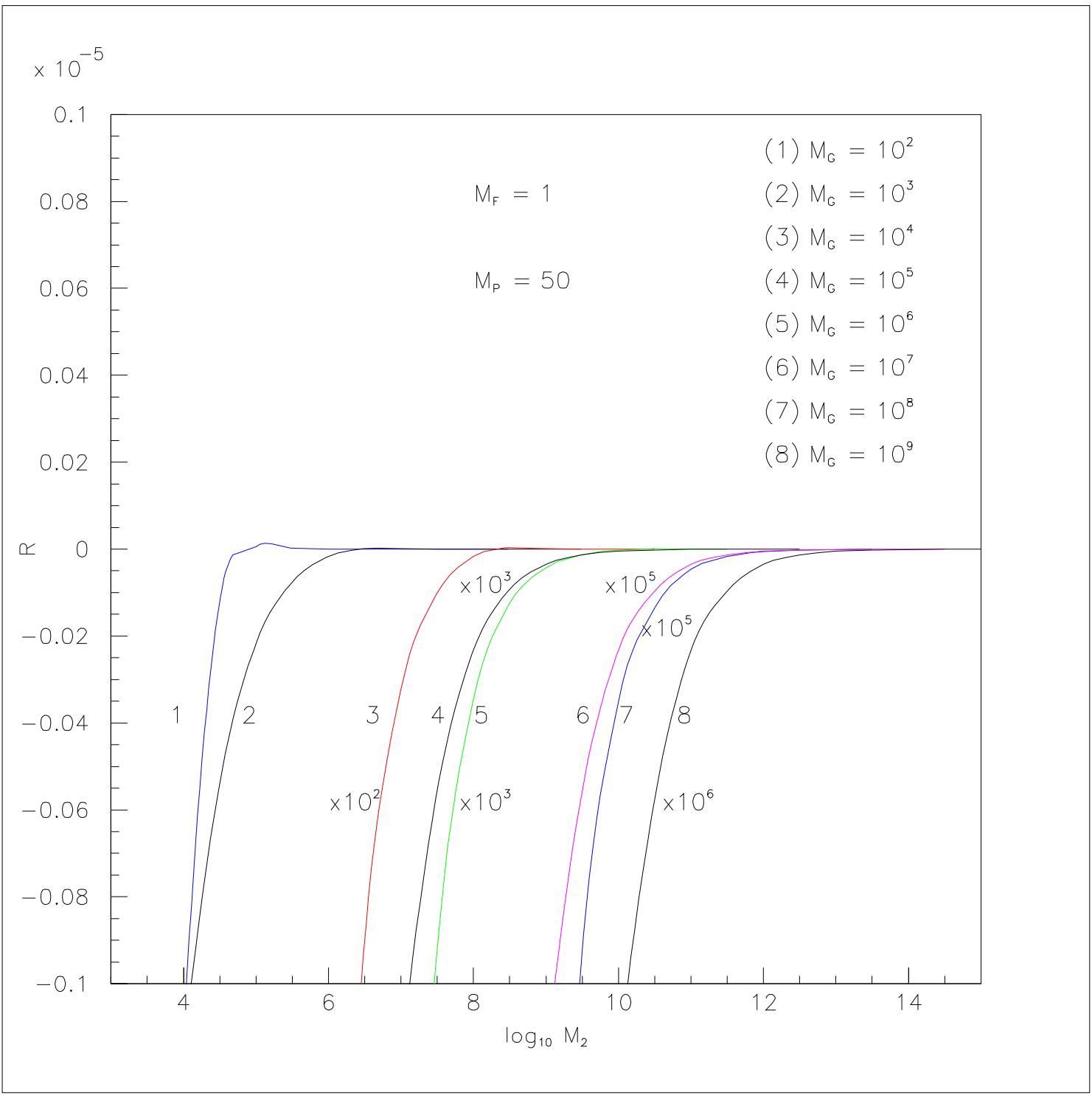
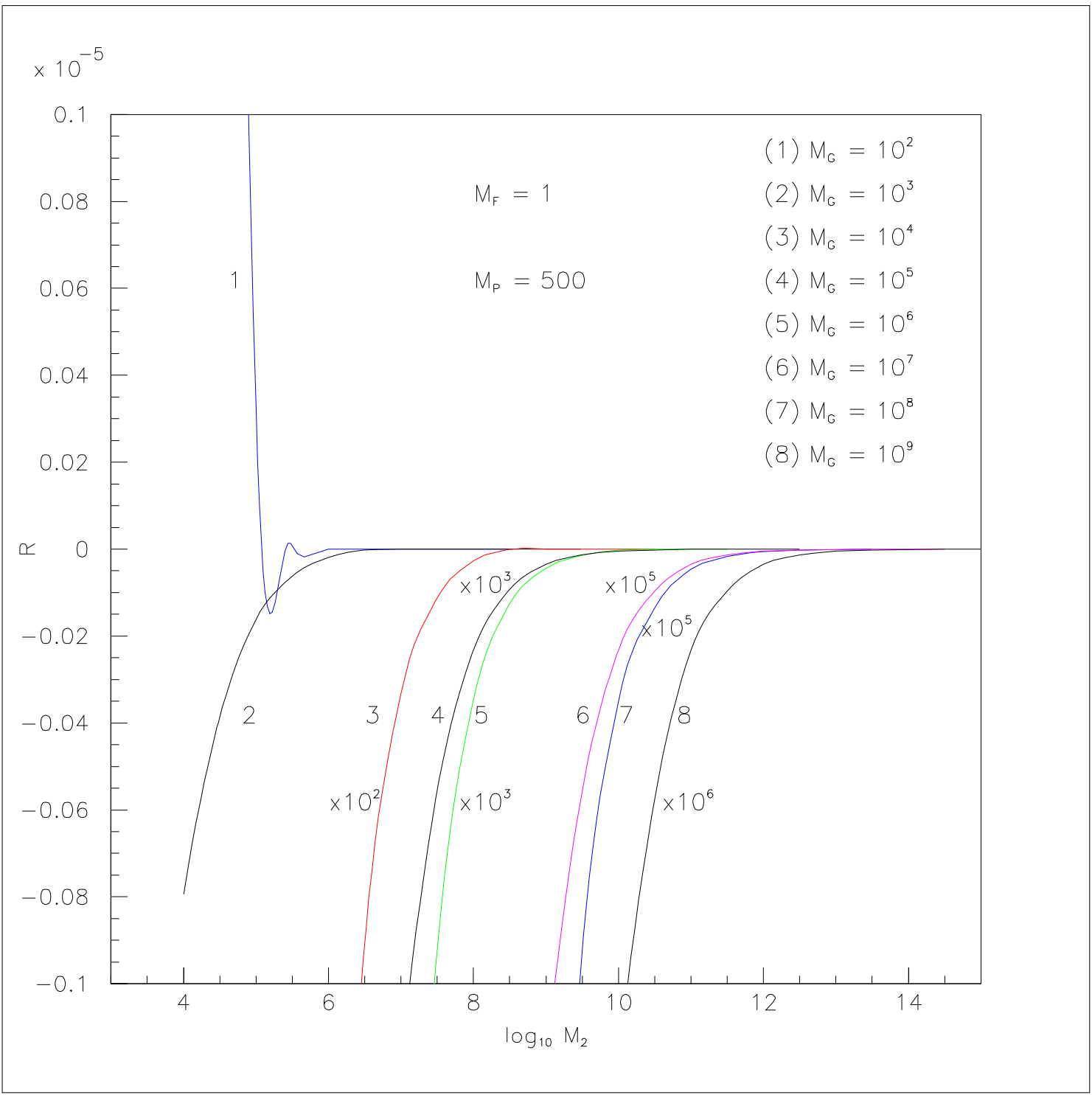
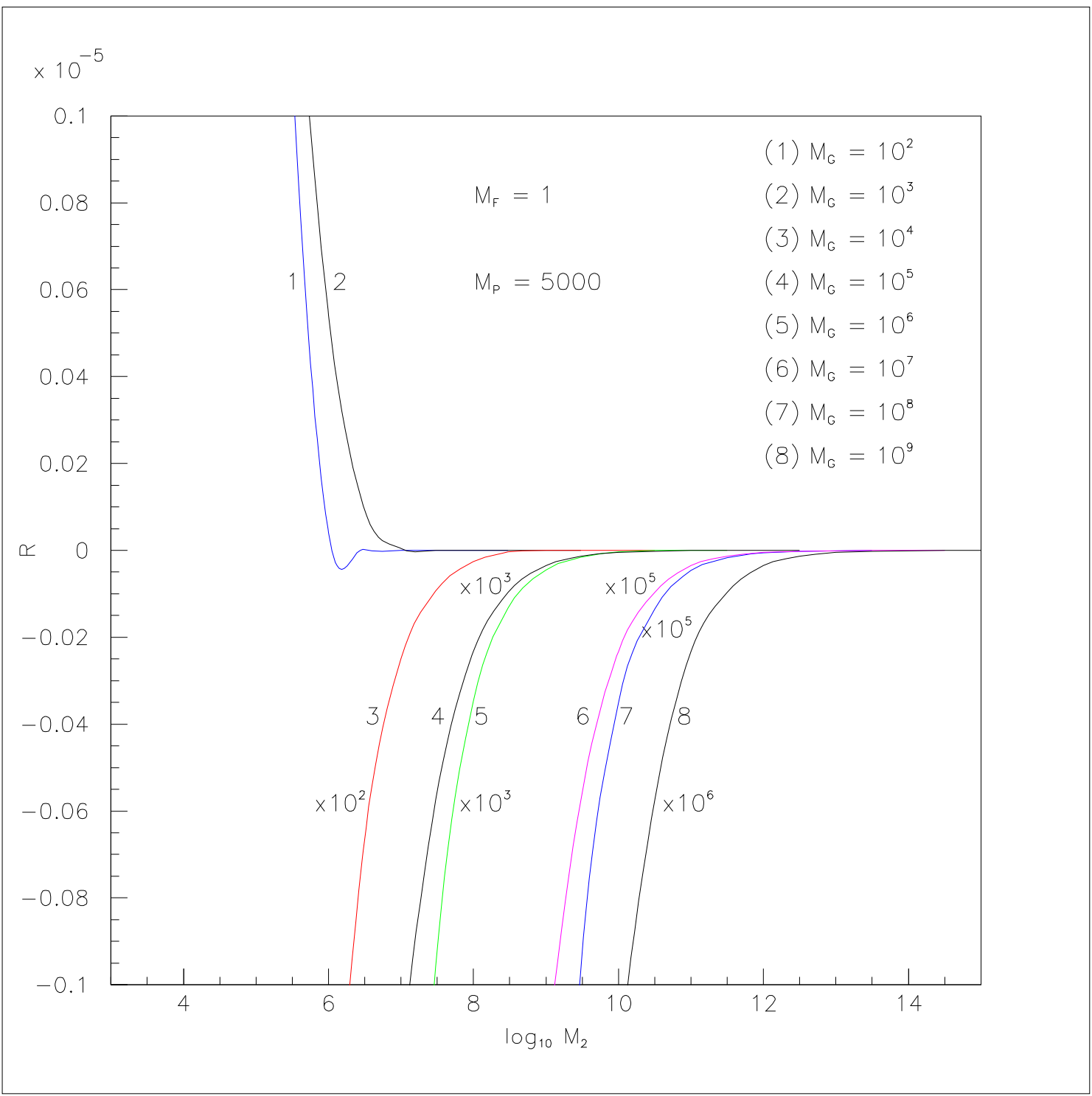
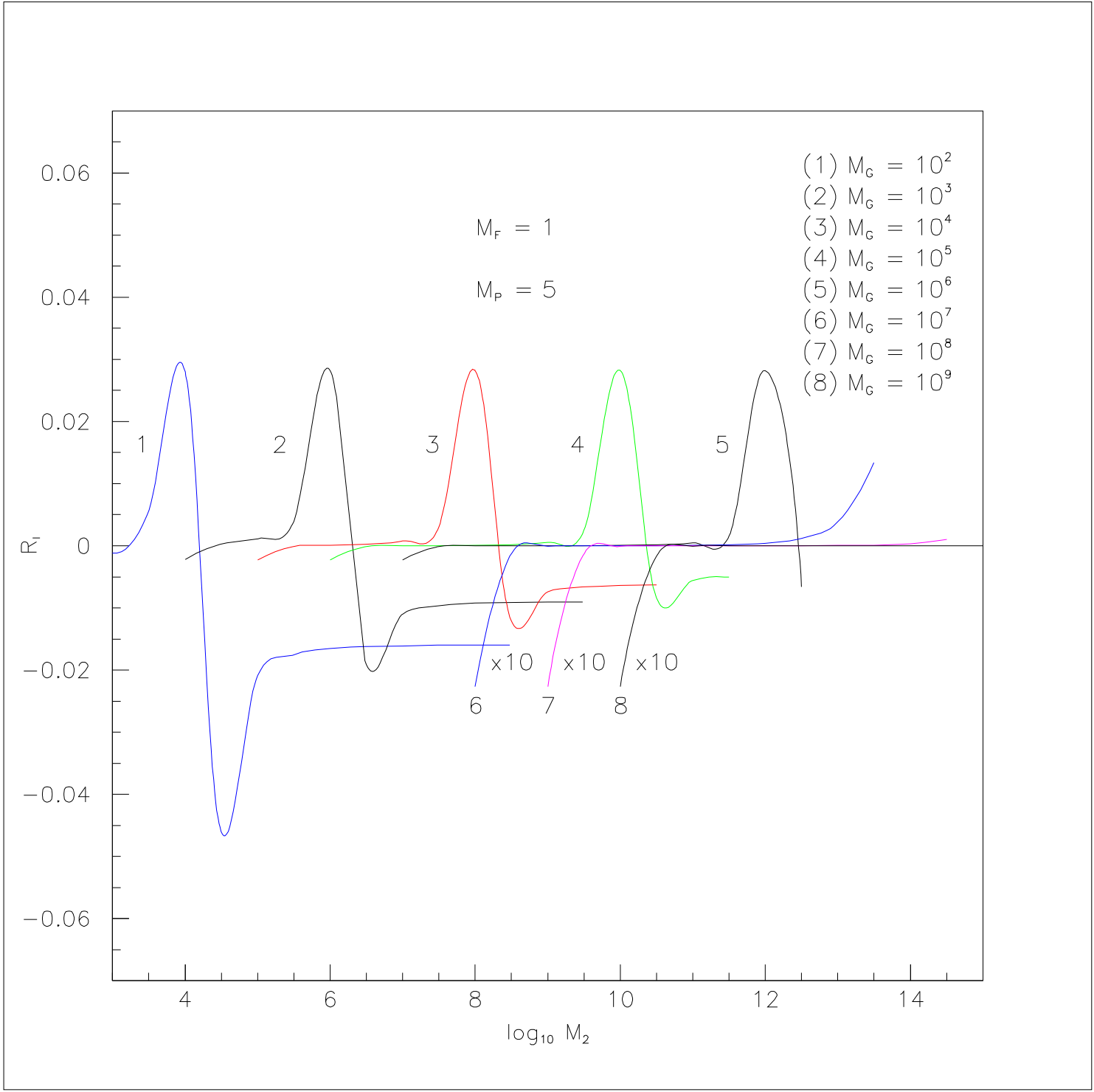
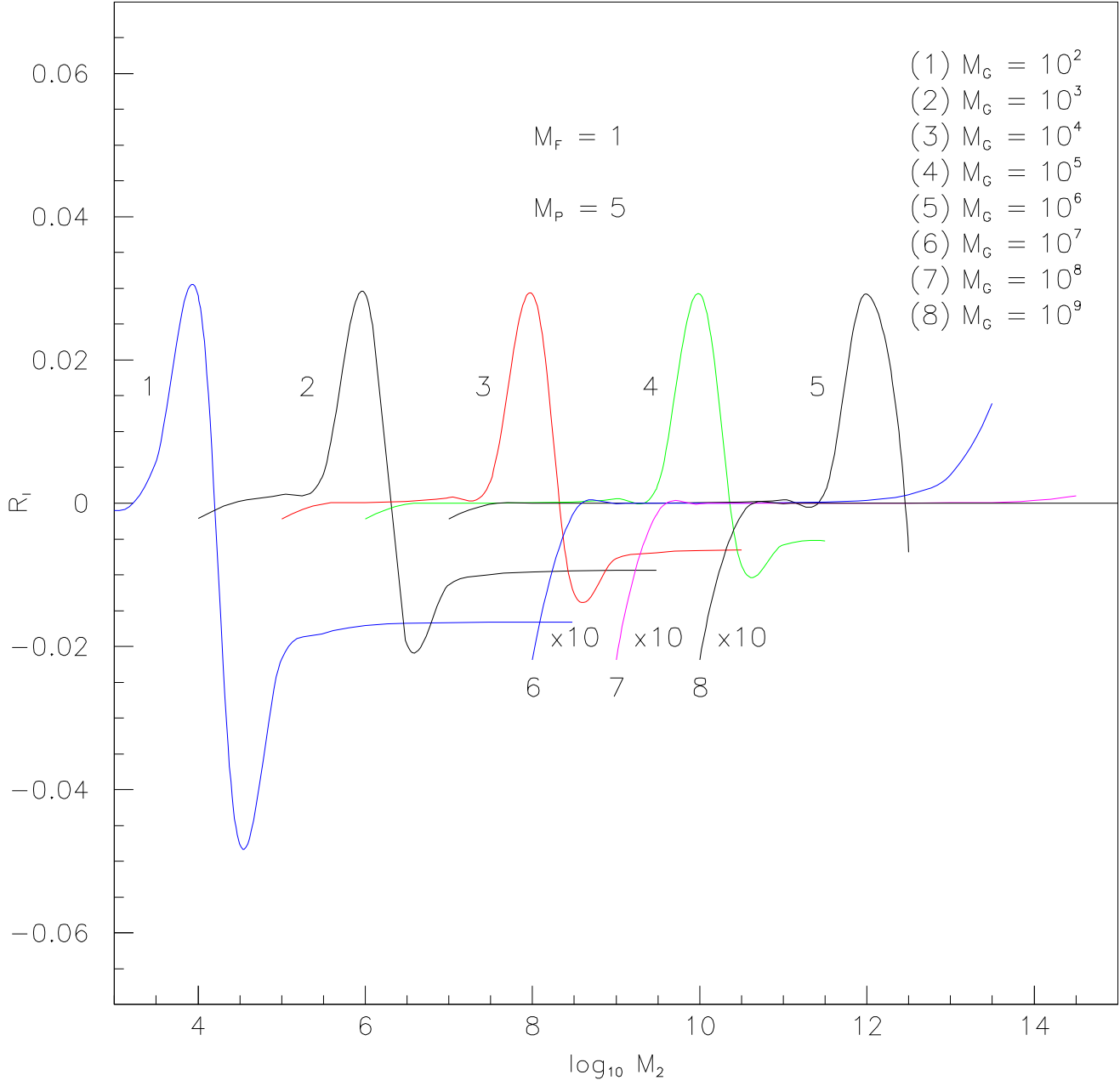
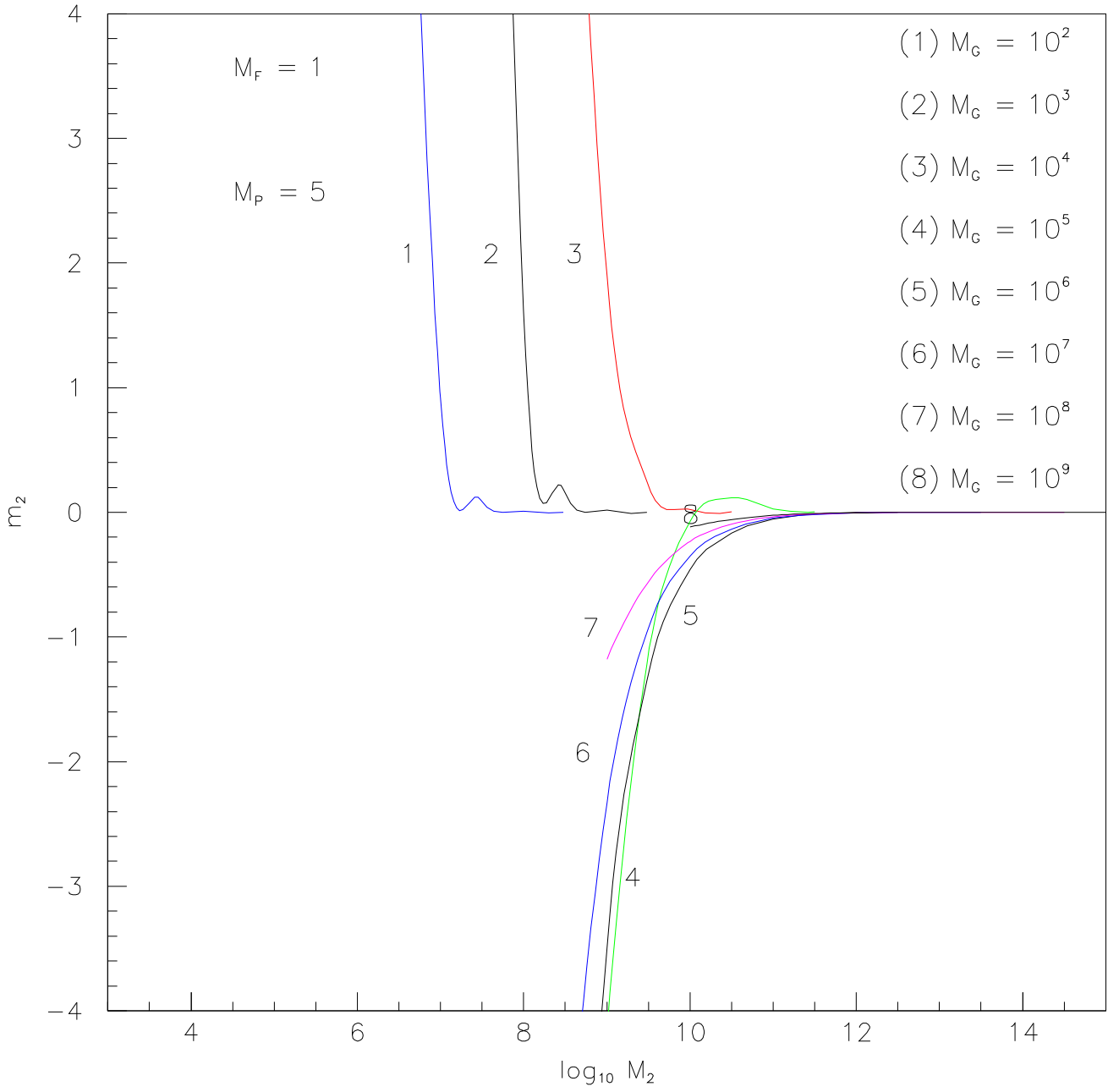
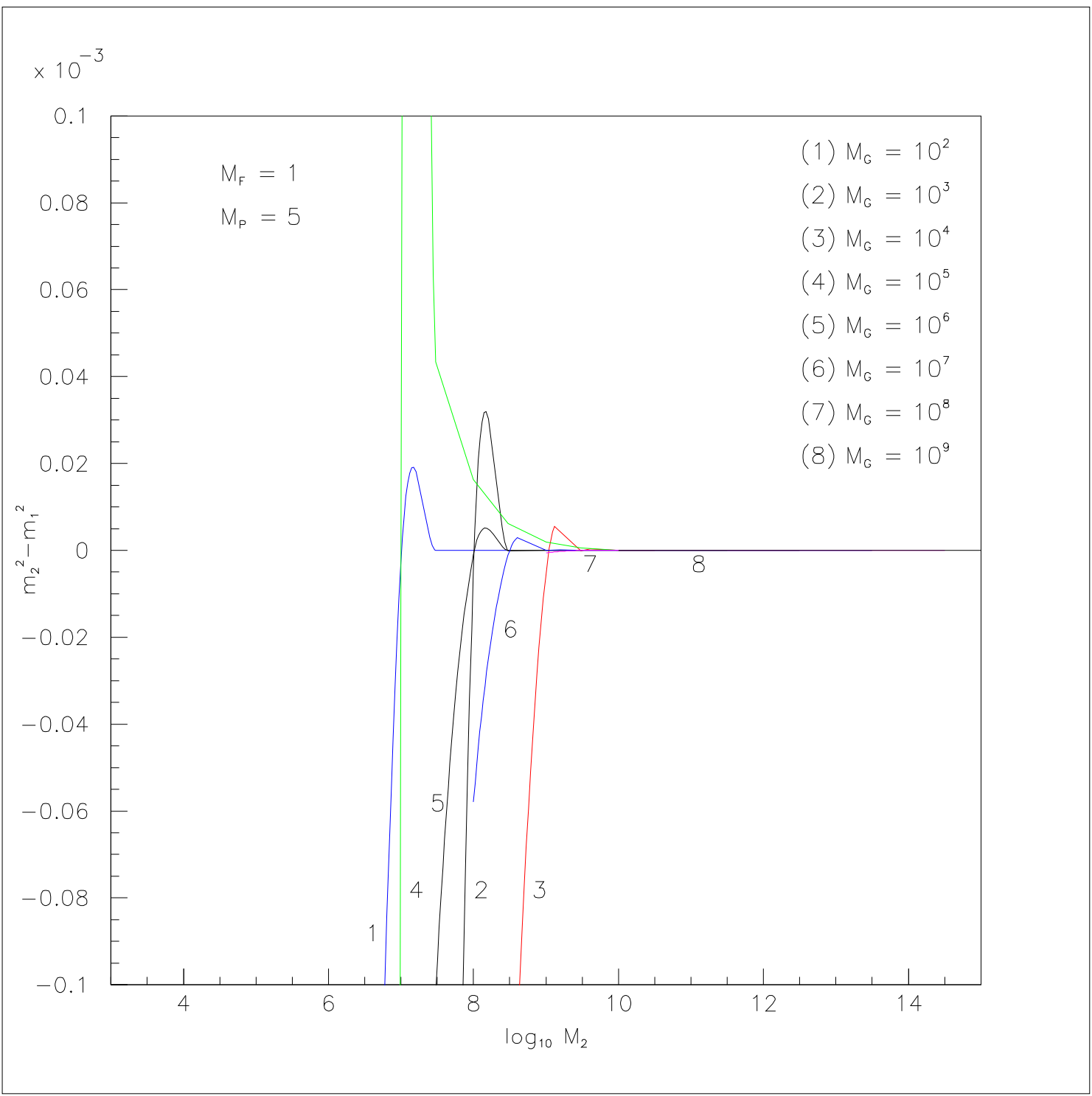
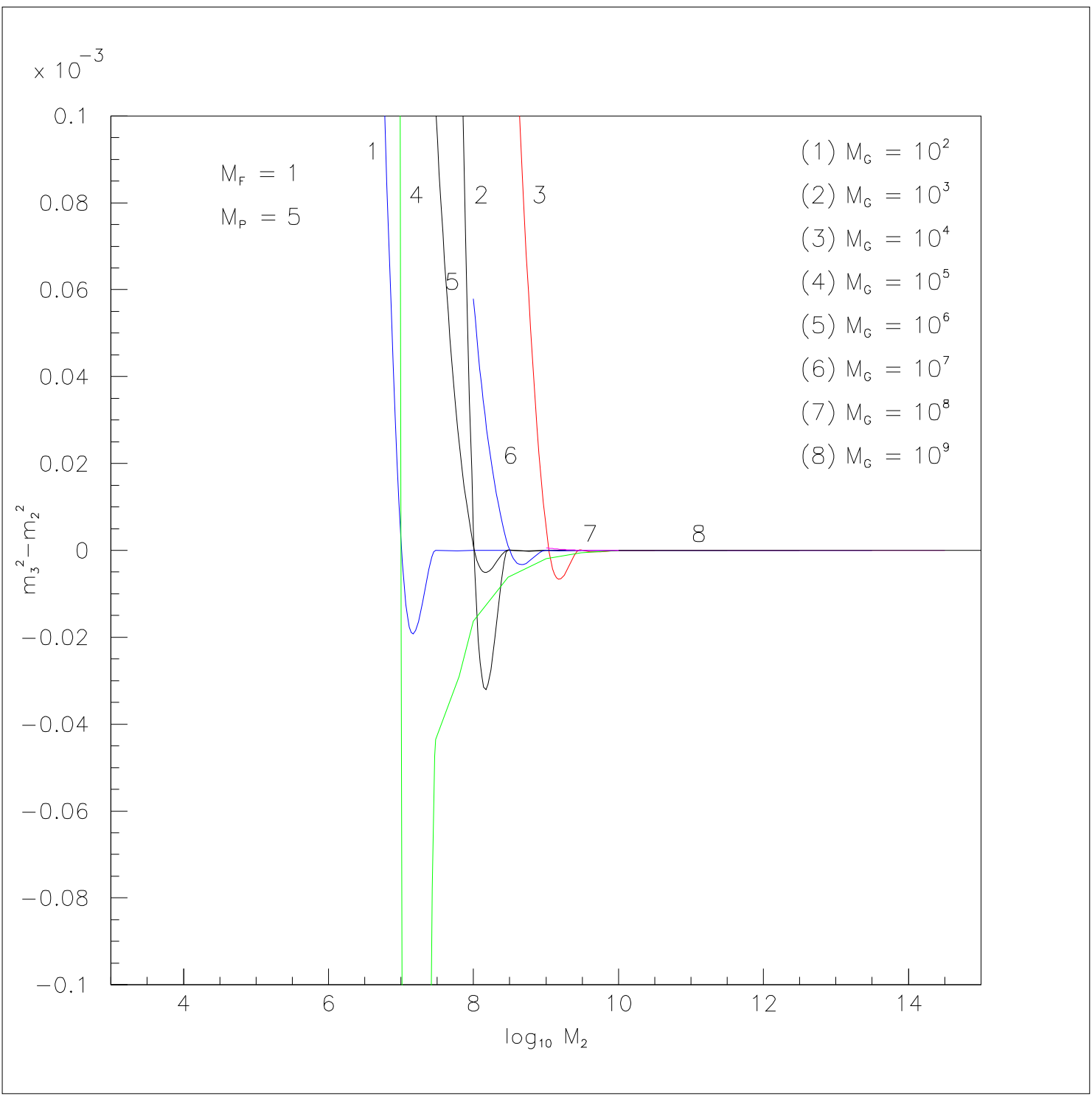
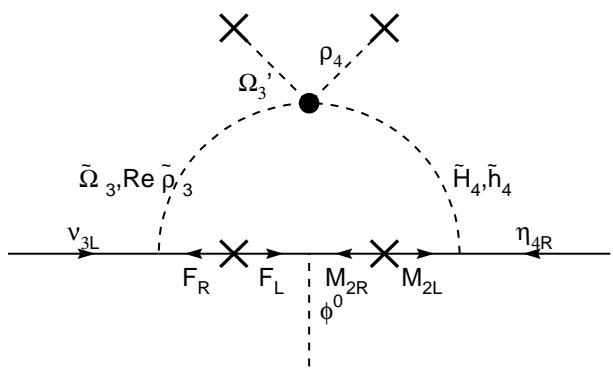
\end{document}